\documentclass[preprint,5p,times,authoryear]{elsarticle}
\usepackage{natbib}
\usepackage{amsmath} 
\usepackage{amsfonts} 
\usepackage{bm} 
\usepackage{calc} 
\usepackage{float}
\usepackage{subcaption} 
\usepackage{hyperref} 

\usepackage{svg}
\usepackage{siunitx}
\usepackage{amssymb,moreverb,pgfplots,tikz}
\pgfplotsset{compat=1.12}
\usepackage{rotating}
\usepackage{enumitem}
\usepackage{cleveref}

\usepackage{algorithm}
\usepackage{algorithmicx}
\usepackage{algpseudocode}

\definecolor{mycolor1}{rgb}{0.00000,0.44700,0.74100}%
\definecolor{mycolor2}{rgb}{0.92900,0.69400,0.12500}%
\definecolor{MatlabBlue}{rgb}    {0     , 0.4470, 0.7410}
\definecolor{MatlabRed}{rgb}     {0.8500, 0.3250, 0.0980}
\definecolor{MatlabYellow}{rgb}  {0.9290, 0.6940, 0.1250}
\definecolor{MatlabPurple}{rgb}  {0.4940, 0.1840, 0.5560}
\definecolor{MatlabGreen}{rgb}   {0.4660, 0.6740, 0.1880}
\definecolor{MatlabBabyBlue}{rgb}{0.3010, 0.7450, 0.9330}
\definecolor{MatlabGray}{rgb}{0.5, 0.5, 0.5}
\definecolor{MatlabLightGray}{rgb}{0.75, 0.75, 0.75}
\definecolor{MatlabBlack}{rgb}{0, 0, 0}

\definecolor{MatlabLightGray4}{rgb}{0.875, 0.875, 0.875}
\definecolor{MatlabLightGray3}{rgb}{0.85, 0.85, 0.85}
\definecolor{MatlabLightGray2}{rgb}{0.775, 0.775, 0.775}
\definecolor{MatlabLightGray1}{rgb}{0.7, 0.7, 0.7}
\definecolor{MatlabGray30}{rgb}{0.3, 0.3, 0.3}
\definecolor{MatlabGray40}{rgb}{0.4, 0.4, 0.4}
\definecolor{MatlabGray50}{rgb}{0.5, 0.5, 0.5}
\definecolor{MatlabGray60}{rgb}{0.6, 0.6, 0.6}
\definecolor{MatlabGray70}{rgb}{0.7, 0.7, 0.7}
\definecolor{MatlabGray80}{rgb}{0.8, 0.8, 0.8}
\definecolor{MatlabGray90}{rgb}{0.9, 0.9, 0.9}

\definecolor{Red}{rgb}{1 0 0}
\definecolor{Black}{rgb}{0 0 0}

\definecolor{myblue}{rgb}{0 0 0}

\newcommand{\tikzline}[1]{(\protect\tikz[baseline=-0.6ex,x=1pt,y=1pt]{ \protect\draw[#1,thick] [-] (0,0) -- (10,0);})}
\newcommand{\tikzdashedline}[1]{(\protect\tikz[baseline=-0.6ex,x=0.9pt,y=1pt]{ \protect\draw[#1,thick,dashed] [-] (0,0) -- (10,0);})}

\DeclareRobustCommand\encircle[1]{%
\tikz[baseline=(X.base)] 
   \node (X) [draw, shape=circle, inner sep=0] {\strut #1};}

\newcommand{\tikzsquare}[1]{ (\protect\tikz[baseline=0ex,x=1pt,y=1pt]{\protect\draw[fill=#1,draw=none,rounded corners= 1pt] (0,0) rectangle (5.5,5.5);})}

\hypersetup{pdfauthor={Name}} 

\usepackage{amsthm}

\newtheorem{assumption}{Assumption}

\newtheorem{lemma}{Lemma}

\newtheorem{proposition}{Proposition}
\newtheorem{theorem}{Theorem}
\newtheorem{example}{Example}[section]
\newtheorem{remark}{Remark}[section]

\usepackage{graphicx}

\makeatletter
\def\ps@pprintTitle{%
  \let\@oddhead\@empty
  \let\@evenhead\@empty
  \let\@oddfoot\@empty
  \let\@evenfoot\@oddfoot
}
\makeatother

\usepackage{eso-pic} 
\AddToShipoutPictureBG*{%
\AtPageUpperLeft{%
\setlength\unitlength{1in}%
\hspace*{\dimexpr0.5\paperwidth\relax} 
\makebox(0,-1.75)[c]{ 
\begin{tabular}{c c} 
Koen Classens \emph{et al.},

Recursive Identification of Structured Systems: An Instrumental-Variable Approach\\  Applied to Mechanical Systems, 

{\em to appear in European Journal of Control}, 

uploaded to ArXiv \today \\ 
\end{tabular}}}}

\begin{document}
    
\begin{titlepage}
    \title{Recursive Identification of Structured Systems:\\ An Instrumental-Variable Approach Applied to Mechanical Systems}
    
    \author[1]{Koen Classens\corref{cor1}}
    \ead{k.h.j.classens@tue.nl}
    \author[1]{Rodrigo A. Gonz\'alez}
    \ead{r.a.gonzalez@tue.nl}
    \author[1,2]{Tom Oomen}
    \ead{t.a.e.oomen@tue.nl}
    
    \cortext[cor1]{Corresponding author}
   
    \address[1]{Eindhoven University of Technology, Department of Mechanical Engineering, Control Systems Technology, Eindhoven, The Netherlands}
    \address[2]{Delft University of Technology, Department of Mechanical Engineering, Delft Center for Systems and Control, Delft, The Netherlands}


\begin{abstract}
    Online system identification algorithms are widely used for monitoring, diagnostics and control by continuously adapting to time-varying dynamics. Typically, these algorithms consider a model structure that lacks parsimony and offers limited physical interpretability. The objective of this paper is to develop a real-time parameter estimation algorithm aimed at identifying time-varying dynamics within an interpretable model structure. An additive model structure is adopted for this purpose, which offers enhanced parsimony and is shown to be particularly suitable for mechanical systems. The proposed approach integrates the recursive simplified refined instrumental variable method with block-coordinate descent to minimize an exponentially-weighted output error cost function. This novel recursive identification method delivers parametric continuous-time additive models and is applicable in both open-loop and closed-loop controlled systems. Its efficacy is shown using numerical simulations and is further validated using experimental data to detect the time-varying resonance dynamics of a flexible beam system. These results demonstrate the effectiveness of the proposed approach for online and interpretable estimation for advanced monitoring and control applications.
\end{abstract}


\end{titlepage}
    
    \maketitle

\section{Introduction}
System identification involves obtaining mathematical models of complex phenomena from measured data, with wide-ranging applications across diverse fields \citep{soderstrom1989system,ljung1998system}. The development of accurate models enables the simulation, analysis, and prediction of complex system behavior. In control engineering, models are the key to designing effective control strategies, while its predictive capabilities support decision-making and planning.

Online system identification plays a significant role in dynamic environments where systems are prone to change over time \citep{ljung1983theory,young2011recursive}. Unlike offline identification, which relies on a batch of data, online system identification continuously updates models in real-time as new data becomes available. Online system identification allows for quick adjustments to evolving conditions, which can enhance model accuracy, improve performance, and facilitate effective maintenance of various industrial process \citep{classens2023Opportunities}. Consequently, it is instrumental in maintaining optimal system performance, enabling timely fault detection and diagnosis, and supporting adaptive control in dynamic unpredictable settings.

Estimation of continuous-time (CT) dynamical models of industrial processes offers advantages over discrete-time (DT) models. Continuous-time models can provide a more accurate representation of the underlying dynamics of physical systems, since they allow for the direct incorporation of \textit{a priori} knowledge such as the relative degree of the physical systems they model \citep{garnier2014advantages}. Incorporating prior knowledge in the form of interpretable physics has gained increasing importance in the fields of system identification, control, monitoring, and machine learning. In the context of linear system identification, incorporating relative degree information is particularly useful when the time evolution of the signals is naturally continuous, as seen in mechanical systems \citep{gawronski2004advanced}, where impulse response discontinuities are typically absent due to the double integration relationship between force and position.

Although a single non-additive transfer function is commonly employed in linear system identification, practical applications in flexible motion systems \citep{oomen2018advanced,voorhoeve2020identifying}, and vibration analysis are more effectively conceptualized as a sum of transfer functions with distinct denominators. For example, mechanical systems are interpreted more naturally in a modal form, where each submodel represents a different resonant mode \citep{gawronski2004advanced}. Additive model parametrizations bring benefits such as enhanced physical insight for fault diagnosis \citep{classens2022fault} and improved numerical conditioning for high-order systems \citep{gonzalez2023identification}. These parameterizations are also explored in statistics \citep{hastie1986generalized} and econometrics \citep{opsomer1999root,hardle2004bootstrap}, offering increased model flexibility and the ability to decentralize analysis for optimization and control purposes. Despite contributions in nonlinear discrete-time finite-impulse response and generalized Hammerstein model estimation \citep{bai2005identification,bai2008identification}, the utilization of additive model structures in system identification has been limited.

The identification of additive models has recently gained increasing attention. An algorithm for offline estimation of continuous-time additive models in single-input single-output (SISO) systems, applicable to both open and closed-loop configurations, has been introduced \citep{gonzalez2023identification}. This approach is based on the simplified refined instrumental variable method, SRIVC \citep{young1980refined}, which consists of iteratively computing instrumental variable estimates of the system parameters. Alternatively, a block coordinate algorithm is utilized for identifying these systems under this structure with a tailored block-coordinate descent version of the SRIVC method \citep{gonzalez2023parsimonious}.

In contrast to these offline algorithms, many recursive algorithms have been developed by parameterizing the model as a single non-additive transfer function \citep{ljung1983theory,young2011recursive}. These online methods are typically based on a forgetting factor, or in state estimation mechanisms analogous to the Kalman filter. Most methods are categorized as modifications of offline identification methods, (non-)linear filtering methods, stochastic approximation methods, and pseudolinear regression-based methods \citep{ljung1983theory}. Methods for time-varying systems, such as \citet{padilla2017recursive,padilla2019identification} explore open and closed-loop approaches based on prefiltering the input and output derivatives with ad-hoc prefilters, similar to the state-variable filter and refined instrumental variable methods for linear and time-invariant systems \citep{young1980refined,garnier2008book}.

Although current recursive methods can effectively track time-varying systems, these yield non-additive transfer function models with limited interpretability, as the estimated parameters may not directly correspond to physical system quantities. In this paper, additive linear time-varying systems are considered for which the parameter variation is slow compared to the system dynamics. To track these systems, a recursive counterpart is developed of the algorithm in \citep{gonzalez2023parsimonious}, where instrumental variable iterations are performed for each submodel until convergence, aligning with the philosophy of the block coordinate descent approach for non-convex optimization \citep{luenberger2008linear}. A comprehensive identification method for modeling additive linear time-varying continuous-time systems is proposed for both open and closed-loop settings. The method is referred to as recursive SRIVC for additive systems, or abbreviated as additive RSRIVC.

In summary, the main contributions are summarized as follows.
\begin{enumerate}[label={C\arabic*:}]
    \item A condition is provided under which it is preferable to estimate an additive model structure instead of an unfactored transfer function.
    \item A recursive continuous-time system identification algorithm is derived for systems in additive form, for both open and closed-loop systems and its consistency is assessed. 
     \item The proposed method is evaluated through simulations and is experimentally validated on a flexible beam setup with a time-varying resonance mode.
\end{enumerate}

The remainder of this paper is organized as follows. In \Cref{sec:problem} the system setup is presented and the problem is formulated. \Cref{sec:parsim_interpret} elaborates on the benefits of the proposed approach, showcasing motivational examples and demonstrating where an additive decomposition is preferable to an unfactored transfer function. \Cref{sec:recursive} contains the main contribution of this work, namely the derivation of a recursive identification method for continuous-time systems in additive form. The estimator is numerically evaluated in \Cref{sec:simulations}, and validated using a true mechanical system in \Cref{sec:experiment}. Lastly, \Cref{sec:conclusions} concludes this paper.

\section{System setup and problem formulation}
\label{sec:problem}
Initially, the system setup for additive systems in open-loop and closed-loop configuration is outlined, followed by the formulation of the estimation problem.

\subsection{System setup}
Consider the following continuous-time (CT), linear and time-variant (LTV), single-input single-output (SISO) system
\begin{equation}
       \label{eq:system}
    x(t) = G^{*}(p,t) u(t),
\end{equation}
where $G^{*}(p,t) = B^{*}(p,t)/A^{*}(p,t)$, with $p$ being the Heaviside operator (i.e., $px(t)=\frac{\textnormal{d}x(t)}{\textnormal{d}t}$). The terms $A^{*}(p,t)$ and $B^{*}(p,t)$ are polynomials in $p$ with time-varying coefficients of the form
\begin{align}
\label{eq:apt}
    A^{*}(p,t) &= a^{*}_n(t)p^n + a^{*}_{n-1}(t)p^{n-1} + \cdots + a^{*}_1(t)p + 1, \\
\label{eq:bpt}
    B^{*}(p,t) &= b^{*}_m(t)p^m + b^{*}_{m-1}(t)p^{m-1} + \cdots + b^{*}_1(t)p + b^{*}_0(t),
\end{align}
with $a_{n}^{*} \neq 0$ and $n \geq m$. These polynomials are assumed to be coprime, i.e., for any fixed $t$ along the parameter trajectory, they do not share roots in $p$.

The system in \eqref{eq:system} is represented in its unfactored function form. Any non-additive transfer function that represents \eqref{eq:system} can also be described as a sum of time-variant systems of lower order. Such decomposition is called an additive form of \eqref{eq:system}, which is described by
\begin{subequations}
    \label{eq:systemadditive}
\begin{align}
    x_i(t) &= G_i^{*}(p,t) u(t), \\
    x(t) &= \sum_{i=1}^{K} x_i(t),
\end{align}    
\end{subequations}
where $G_i^{*}(p,t) =B_i^{*}(p,t)/A_i^{*}(p,t)$, and $A^{*}_i(p,t)$ and $B^{*}_i(p,t)$ have the same form as \eqref{eq:apt} and \eqref{eq:bpt}, but have orders $n_i$ and $m_i$ with $a_{i,n_{i}}^{*} \neq 0$, $n_i \geq m_i$, and $\sum_{i=1}^{K}n_i = n$. That is,
\begin{align}
\label{eq:apt2}
    A^{*}_i(p,t) &= a^{*}_{i,n_{i}}(t) p^{n_{i}} + a^{*}_{i,n_{i}-1}(t) p^{n_{i}-1} + \cdots + a^{*}_{i,1}(t) p + 1, \\
\label{eq:bpt2}
    \begin{split}
B^{*}_{i}(p,t) &= b^{*}_{i,m_{i}}(t) p^{m_{i}} + b^{*}_{i,m_{i}-1}(t) p^{m_{i}-1} + \cdots + b^{*}_{i,1}(t) p \\ &+ b^{*}_{i,0}(t).
\end{split}
\end{align}
Without loss of generality, it is assumed that the polynomials $A^{*}_i(p,t)$ are jointly coprime and anti-monic, i.e., their constant coefficient is fixed to one. In addition, for each separate submodel to be identifiable, the system is parameterized such that at most one subsystem $G^{*}_i(p,t)$ is biproper. An approach to determine an additive decomposition of \eqref{eq:system} is to calculate the partial fraction decomposition of $B^{*}(p,t)/A^{*}(p,t)$.

\begin{figure}[]
  \centering
  \includegraphics[trim={0mm 0mm 0mm 0mm},clip]{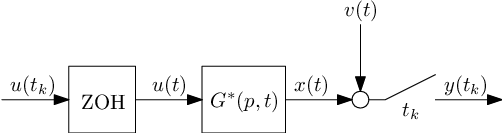}\vspace{5.75mm}
	\caption{Block diagram for the open-loop setting studied in this paper.}
  \label{fig:ol_setup}
\end{figure}

The polynomials $A_i^*(p,t)$ and $B_i^*(p,t)$ are jointly described by the parameter vector 
\begin{equation}
\begin{split}
\label{ctparametervector}
	\theta_i^*(t) = &[
		a_{i,1}^*(t) \:\:\:\: a_{i,2}^*(t) \:\:\:\: \dots \:\:\:\: a_{i,n_i}^*(t) \\ & \qquad b_{i,0}^*(t) \:\:\:\: b_{i,1}^*(t) \:\:\:\: \dots \:\:\:\: b_{i,m_i}^*(t)	
  ]^\top,
\end{split}
\end{equation}
and
\begin{equation}
\label{eq:betastar}
\beta^*(t):= \begin{bmatrix}
\theta_1^{*\top}(t) & \theta_2^{*\top}(t) & \dots & \theta_{K}^{*\top}(t)
\end{bmatrix}^\top.
\end{equation}
A noisy output measurement is retrieved at equidistantly-spaced time instants $t=t_k$, see \Cref{fig:ol_setup}, i.e.,
\begin{equation}
    y(t_k)=x(t_k) + v(t_k),
      \label{eq:output}
\end{equation}
where $v$ is a stochastic process of zero mean and finite variance $\sigma^{2}$.

\begin{figure}[]
  \centering
  \includegraphics[trim={0mm 0mm 0mm 0mm},clip]{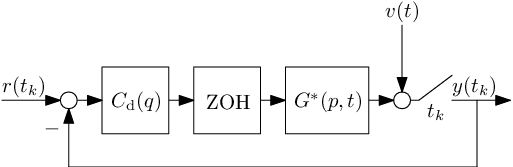}
	\caption{Block diagram for the closed-loop setting studied in this paper.}
  \label{fig:cl_setup}
\end{figure}

If $G^*(p,t)$ is in closed loop, see \Cref{fig:cl_setup}, it is assumed that the system is driven by a reference signal $r(t_{k})$ and the control input is determined through the control law
\begin{equation}
    u(t_k) = C_{\mathrm{d}}(q)(r(t_k)-y(t_k)),
          \label{eq:CLinput}
\end{equation}
where $C_{\mathrm{d}}(q)$ denotes the (discrete-time) controller transfer function, and $q$ denotes the forward shift operator. It is assumed that the input sequence $u(t_k)$ is interpolated via a zero-order hold prior to being used to excite the system. Substitution of the output $y(t_k) = G_\textnormal{d}^*(q,t_k) u(t_k) + v(t_k)$ gives the input described as 
\begin{equation}
    u(t_k) = \tilde{r}(t_k)-\tilde{v}(t_k),
          \label{eq:CLinput2}
\end{equation}
where
\begin{equation}
	\label{eq:rtk}
	\tilde{r}(t_k):=S_{uo}^*(q,t_k) r(t_k), \quad 	\tilde{v}(t_k):=S_{uo}^*(q,t_k) v(t_k),
\end{equation}
with $S_{uo}^*(q,t_k)$ the sensitivity function $S_{uo}^*(q,t_k)=$ $C_{\textnormal{d}}(q)/[1+G_\textnormal{d}^*(q,t_k)C_\textnormal{d}(q)]$, and $G_\textnormal{d}^*(q,t_k)$ being the zero-order-hold equivalent discrete-time system of $G^*(p,t)=\sum_{i=1}^K G_i^*(p,t)$. Explicit expressions for computing $G_\textnormal{d}^*(q,t_k)$ are obtained when changes in the parameters in $A^*(p,t)$ and $B^*(p,t)$ occur only at the sampling time instants $t_k$, see \citet[Chap. 6.3.1]{toth2010modeling} for more details.

\subsection{Problem formulation}
\label{subsec:problem_formulation}
Consider the CT LTV system with input $u(t)$ and output $y(t)$ in \eqref{eq:systemadditive} and \eqref{eq:output}. The data-generating system is described by \eqref{eq:output} in the open-loop setting. In case the system is closed-loop controlled, the data-generating system additionally includes \eqref{eq:CLinput} and the setpoint $r(t)$.

It is assumed that the plant is represented by the model set
\begin{subequations}
    \label{eq:systemadditive_modelset}
\begin{align}
    x_i(t) &= G_i(p,t,\theta_{i})u(t), \\
    x(t) &= \sum_{i=1}^{K} x_i(t), \\
    y(t_{k}) &= x(t_{k}) + v(t_{k}).
\end{align}    
\end{subequations}
Here $G_i(p,t,\theta_i) = B_i(p,t,\theta_{i})/A_i(p,t,\theta_{i})$, where the polynomials $A_i(p,t,\theta_{i})$ and $B_i(p,t,\theta_{i})$ have the same structure as \eqref{eq:apt2} and \eqref{eq:bpt2} with $\theta_{i} (t) \in \mathbb{R}^{n_{\theta_{i}}}$, and $n_{\theta_{i}} = n_{i} + m_{i} + 1$. The estimate of the total additive form is described by $\beta(t) \in \mathbb{R}^{n_{\beta}}$, which has the same structure as \eqref{eq:betastar} and where $n_{\beta} = \sum_{i=1}^K (n_i+m_i) + K$.

The considered problem is to recursively estimate $\beta(t)$ which characterizes the model structure given by \eqref{eq:systemadditive_modelset}, based on the input and output data $\{u(t_k),y(t_k)\}_{k=1}^N$. Here, $N$ is the number of samples available at time $t_{N}$ for the open-loop scenario, while $\{r(t_k)\}_{k=1}^N$ is also known in the closed-loop setting.

Next, the primary benefits of additive parameterizations are presented, supported by multiple motivating examples.

\section{Parsimony and interpretability in continuous-time system identification}
\label{sec:parsim_interpret}
The estimation in an additive form has several benefits including the following.
\begin{enumerate}
    \item Direct continuous-time identification methods may suffer from a lack of parsimony when the sum of transfer functions of particular relative degrees is estimated \citep{gonzalez2023parsimonious}.
    \item The parameters in an additive structure may be more interpretable compared to an unfactored transfer function.
    \item Typically, derivatives of the input and output data are required for recursive estimation. Considering an additive structure can significantly reduce the order of the derivatives to be estimated from sampled data.
\end{enumerate}

First, the advantages with respect to parsimony and interpretability are described. Later, in \Cref{subsec:RecursiveSRIVC}, \Cref{rem:derivativeorder}, the additional advantage regarding the derivative order is highlighted.

Direct continuous-time identification approaches such as the SRIVC method may suffer from a lack of parsimony when identifying a system that consists of the sum of transfer functions. The following proposition quantifies the additional parameters being estimated when opting to estimate \eqref{eq:system} with a model structure $G(p)=B(p)/A(p)$ of relative degree $r$ instead of separately estimating the parameters for each transfer function $B_i(p)/A_i(p)$.

\begin{proposition} \citep{gonzalez2023parsimonious}
	\label{prop:prop1}
	Consider the system in \eqref{eq:system}, and the model structure $G(p)=\sum_{i=1}^K B_i(p)/A_i(p)$. Opting for the model structure $G(p)=B(p)/A(p)$ during identification (with the minimal relative degree $r$ that encompasses the true system) results in a lack of parsimony for the latter model structure if and only if
	\begin{equation}
		\label{eq:excess}
		\sum_{i=1}^{K} r_i - r> K-1,
	\end{equation}
	where $r_i=n_i-m_i$ is the relative degree of $B_i(p)/A_i(p)$. The surplus in \eqref{eq:excess} is the number of additional parameters considered by the model structure $G(p)=B(p)/A(p)$. 
\end{proposition}

An immediate implication derived from Proposition \ref{prop:prop1} is that when $r=0$ or $r=1$, the model structure $G(p)=B(p)/A(p)$ suffers from a lack of parsimony if a transfer function $G_i^*(p)$ exists in its partial fraction expansion \eqref{eq:system} with a relative degree greater than one. Two examples of this property are given below.
\begin{example}
	Consider the system
	\begin{equation}
		G^*(p) = \frac{b_{1,0}^*}{a_{1,1}^* p +1} + \frac{b_{2,0}^*}{a_{2,2}^* p^2 +a_{2,1}^* p +1}, \notag 
	\end{equation}
	where the $a^*_{i,j}$ values are strictly positive, and the $b_{i,0}^*$ are non-zero. Only 5 parameters must be estimated if the following model structure is used:
	\begin{equation}
		\label{eq:parsimoniousmodelstructure}
		G(p)=\frac{b_{1,0}}{a_{1,1} p +1} + \frac{b_{2,0}}{a_{2,2} p^2 +a_{2,1} p +1}.
	\end{equation}
Alternatively, if the consideration of this partial fraction decomposition is disregarded and the decision is made to employ the model structure
    \begin{equation}
		G(p)=\frac{b_{2}p^2+b_{1}p+b_{0}}{a_{3} p^3 + a_{2} p^2 + a_{1} p +1} \notag 
	\end{equation}
	with no constraints on the parameter values, then 6 parameters must be estimated. This model structure leads to a lack of parsimony compared to the one in \eqref{eq:parsimoniousmodelstructure}. \hspace*{\fill}  $\triangle$
\end{example}
\begin{example}
	Consider the system
	\begin{equation}
		G^*(p) = \frac{b_{0}^*}{a_{2}^* p^2+a_{1}^* p +1}, \notag 
	\end{equation}
	where $a^*_{1}$ and $a_2^*$ are strictly positive, and $a^*_1>2\sqrt{a_2^*}$. Hence, $G^*(p)$ has two real-valued poles. If the following partial fraction decomposition of $B^*(p)/A^*(p)$ is considered
	\begin{equation}
		G(p)=\frac{b_{1,0}}{a_{1,1} p +1} + \frac{b_{2,0}}{a_{2,1} p +1}, \notag
	\end{equation}
	then 4 parameters must be estimated instead of 3. This additive representation does not directly take into account the relationship between the numerator terms $b_{1,0}$ and $b_{2,0}$, leading to a lack of parsimony. \hspace*{\fill}  $\triangle$
\end{example}

The following example shows a mechanical system in one of its canonical forms. In addition to parsimony for this type of system, the interpretability of the parameters is also examined.
\begin{example} \label{example:mechsys}
	Mechanical systems are typically described as a sum of modes, i.e.,
\begin{equation}
    G^*(p)=\sum_{i=1}^K \frac{b_{i,0}^*}{p^2/{\omega_i^{*}}^2 + 2(\zeta_i^{*}/\omega_i^{*}) p + 1}, \label{eq:mechsys} 
\end{equation}
  where $\omega_{i}^{*} > 0$ and $\zeta_i^{*} > 0$ denote the natural frequency and the damping coefficient of the $i$th subsystem \citep{gawronski2004advanced}. These systems are described by $3 K$ parameters if the following model structure is used:
	\begin{equation}
		\label{eq:parsimoniousmodelstructure_ex3}
		G(p)=\sum_{i=1}^{K} \frac{b_{i,0}}{a_{i,2} p^2 +a_{i,1} p +1}.
	\end{equation}  
	On the contrary, if the partial fraction decomposition is overlooked and the decision is made to adopt the widely utilized model structure of order $n$ and relative degree two
\begin{equation}
	\label{eq:nonparsimoniousmodelstructure_ex3}
    G(p)=\frac{b_{n-2}p^{n-2} + b_{n-3}p^{n-3} + \ldots + b_{0}}{a_{n} p^{n} + a_{n-1} p^{n-1} \ldots + a_{1} p +1}, 
\end{equation} 
with $n = 2K$ and no constraints on the parameter values, then a total of $4K-1$ parameters must be estimated. This model structure leads to a lack of parsimony compared to the one in \eqref{eq:parsimoniousmodelstructure_ex3} if $K>1$, with a further degradation of parsimony for larger values of $K$. Furthermore, the parameters in \eqref{eq:parsimoniousmodelstructure_ex3} are physically interpretable, as they directly relate to the natural frequencies $\omega_i^{*}$ and the damping coefficients $\zeta_i^{*}$. This interpretability is hidden and is lost mostly in \eqref{eq:nonparsimoniousmodelstructure_ex3}.
  \hspace*{\fill}  $\triangle$
\end{example}

Interpretability of the parameters, as illustrated in the latter example, is highly favourable in many engineering applications. For example, in the field of fault diagnosis, this interpretability greatly simplifies the task of fault isolation, as faults often manifest in a specific submodel. Estimating the system in an unfactored form leads to variations in most parameters, while estimating it in an additive form results in variations solely in the directly related parameters.

\section{Recursive estimation of continuous-time systems: A block coordinate descent method}
\label{sec:recursive}
	
This section provides a recursive solution to estimate additive models. A key optimization tool that is considered in the approach is the block-descent method, which is examined in the next subsection. Then, the recursive estimators are presented for both open and closed-loop settings. In addition, a summary of the algorithm is presented, adaptations for marginally stable systems are given, and practical aspects are described.
	
	\subsection{Block coordinate descent}
	\label{subsec:Block_coordinate_descent}
	Towards the goal of computing a recursive estimator for $\beta^{*}$, consider the minimization problem
\begin{equation}
	\label{eq:op}
    \underset{{i=1,\ldots,K}}{\underset{\theta_i \in \Omega_i}{\min}} V_N (\beta),
	\end{equation}	
where $\Omega_i \subset \mathbb{R}^{n_{\theta_i}}$ is a compact set which contains the true parameters of the $i$th subsystem. The cost $V_{N}$, using the $N$ available samples at $t_N$, is the weighted least squares output error cost function
\begin{equation}
	\label{eq:VN}
    V_N (\beta) = \sum_{k=1}^N \alpha_{N-k}  \left[ y(t_k) - \sum_{i=1}^K G_i(p,t,\theta_{i}) u(t_{k}) \right]^{2},
	\end{equation}	
where the notation $G_i(p,t,\theta_{i})u(t_k)$ means that the input $u(t)$ is filtered through the continuous-time transfer function $G_i(p,t,\theta_{i})$ and later evaluated at $t=t_k$. The function $\alpha$ is a non-negative weighting function of the form
	\begin{equation} \label{eq:FF}
	\alpha_k = \lambda^{k}, \quad \lambda\in(0,1].
	\end{equation}
	Note that $\alpha$ satisfies the multiplicative property $\alpha_{k+1} = \lambda \alpha_k$, with $\alpha_0=1$. The parameter $\lambda$ is often referred to as the forgetting factor \citep{ljung1998system}.
	
The approach to solving \eqref{eq:op} consists of iteratively minimizing the cost with respect to ${\theta}_i$ while leaving the other variables fixed. \Cref{alg:algorithm0} describes the general procedure that is solved, typically after every incoming sample, known as \emph{block coordinate descent} \citep{luenberger2008linear}.
\begin{algorithm}
	\caption{Block coordinate descent}
	\begin{algorithmic}[1]
		\State Choose an initial parameter vector $\theta_i^1$ for each $i$
		\For{$l = 1, 2, \dots$}
		\For{$i = 1, \dots, K$}
		\State $\theta_i^{l+1} \gets \arg\min\limits_{\theta_i \in \Omega_i} V_{N}(\theta_1^{l+1}, \dots, \theta_{i-1}^{l+1}, \theta_i, \theta_{i+1}^{l}, \dots, \theta_K^{l})$
		\EndFor
		\EndFor
	\end{algorithmic}
	\label{alg:algorithm0}
\end{algorithm}

Let $\beta^l := [(\theta_1^l)^\top, \dots, (\theta_K^l)^\top]^\top \in \Omega \subset \mathbb{R}^{n_\beta}$, with $\Omega=\prod_{i=1}^{K} \Omega_i$ being the parameter space. Therefore, each iteration of the algorithm (in $l$) can be written as $\beta^{l+1} = \mathcal{A}(\beta^l)$, where $\mathcal{A}\colon \mathbb{R}^{n_\beta} \to \mathbb{R}^{n_\beta}$ is the composite mapping $\mathcal{A} = \mathcal{T} \circ \mathcal{C}^K \circ \mathcal{T} \circ \mathcal{C}^{K-1} \circ \cdots \circ \mathcal{T} \circ \mathcal{C}^1$. This notation is adopted from \citep{luenberger2008linear}, where the block coordinate descent iterations are referred to as $\mathcal{A}$-iterations. Here, the choice function is defined as $\mathcal{C}^i(\beta) := (\beta, i)$, and the optimization function $\mathcal{T}(\beta,i) := (\theta_1, \dots, \theta_{i-1}$, $\overline{\theta}_i,\theta_{i+1}, \dots, \theta_K)$, with $\overline{\theta}_i$ being defined as $\overline{\theta}_i = \arg\min_{\theta_i \in \Omega_i} V_N(\theta_1,$ $\dots, \theta_i, \dots, \theta_K)$. The following result concerns the convergence of \Cref{alg:algorithm0} to a stationary point of the cost \eqref{eq:op}.

\begin{theorem}[Global convergence of Algorithm \ref{alg:algorithm0}]
	\label{thm:thm42}
	If the search along any coordinate direction $\theta_i$ yields a unique minimum point of $V_N$, then the limit of any convergent subsequence of $\{\beta^l\}$ obtained from $\beta^{l+1} = \mathcal{A}(\beta^l)$ belongs to the set of fixed points $\Gamma_N = \{\beta\in \Omega\colon \nabla V_N(\beta) = 0\}$.
\end{theorem}
\begin{proof}
See, e.g., \citet[Section 8.9]{luenberger2008linear}.     
\end{proof}

\begin{remark}
	\label{rem:remarkglobal}
    The result in \Cref{thm:thm42} still holds if the unique minimum assumption is replaced with the requirement that $V_N$ decreases in each coordinate search. In this case, it is sufficient to find a parameter vector $\bar{\theta}_i$ that reduces the cost $V_N$ instead of minimizing it.
\end{remark}	

The block coordinate descent algorithm described in \Cref{alg:algorithm0} requires a way to compute $\mathcal{T}(\beta,i)$ at each iteration, for each $i=1,2,\dots,K$. That is, given the data up until $t_N$, compute
\begin{equation}
    \begin{split}
        \theta_i^{l+1} &= \underset{\theta_i \in \Omega_i}{\arg \min} \sum_{k=1}^N \alpha_{N-k} \bigg[ y(t_k) - \sum_{j=1}^{i-1} G_j(p,t,\theta_j^{l+1}) u(t_k) \\
	   &-\sum_{j=i+1}^{K} G_j(p,t,\theta_j^l) u(t_k) - G_i(p,t,\theta_i) u(t_k) \bigg]^2
    \end{split}
    \label{eq:opt}
\end{equation}
for $i=1,2,\dots, K$, for fixed values of $\{\theta_j^{l+1}\}_{j=1}^{i-1}$ and $\{\theta_j^l\}_{j=i+1}^{K}$. For a sufficiently large compact parameter space $\Omega_i$, the optimization problem in \eqref{eq:opt} becomes a nonlinear least-squares problem that lends itself to iterative solution methods. To this end, define the residual output of each submodel as
\begin{equation}
	\tilde{y}_i(t_k) := y(t_k)- \sum_{j=1}^{i-1} G_{j}(p,t,\theta_j^{l+1}) u(t_k)-\sum_{j=i+1}^{K} G_{j}(p,t,\theta_j^l) u(t_k). \notag
 \label{eq:ytilde}
\end{equation}
Substitution of this residual output in \eqref{eq:opt} gives
\begin{align}
	\theta_i^{l+1} &= \underset{\theta_i \in \Omega_i}{\arg \min} \sum_{k=1}^N \alpha_{N-k} \bigg[ \tilde{y}_i(t_k) - G_i(p,t,\theta_i) u(t_k) \bigg]^{2}.
	\label{eq:opt2}
\end{align}
Next is shown how this is solved through recursive SRIVC.

\subsection{Recursive SRIVC for additive open-loop systems}
\label{subsec:RecursiveSRIVC}
	In the linear and time-invariant case, when solved iteratively, the optimization problem in \eqref{eq:opt2} leads to a local minimum for the complete optimization problem \eqref{eq:op} \citep{gonzalez2023parsimonious}. Each optimization problem in \eqref{eq:opt2} takes the form of minimizing the weighted sum of squares of the residual equal to $\varepsilon_{i}(t_k,\theta_i)=\tilde{y}_i(t_k)-G_i(p,t,\theta_i) u(t_k)$. This pseudolinear regression takes the general form
	\begin{equation}
	\label{eq:pseudolinear}
	\tilde{y}_{i,\textnormal{f}}(t_k,\theta_i) = \varphi_{i,\textnormal{f}}^\top(t_k,\theta_{i})\theta_{i}(t_k) + \varepsilon_{i}(t_k,\theta_{i}).
	\end{equation}
This generic form admits the refined instrumental variable iterations \cite{young1980refined}. Essentially, the refined instrumental variable iterations provide a sequence which, under mild conditions, converges to a point $\bar{\theta}_{i}$ that satisfies the first order optimality condition of the cost in \eqref{eq:opt2}, i.e.,
\begin{equation}
\sum_{k=1}^N \alpha_{N-k} \frac{\partial \varepsilon_{i} (t_k,\theta_{i})}{\partial \theta_{i}}\bigg|_{\theta_{i}=\bar{\theta}_{i}} \varepsilon_{i}(t_k,\bar{\theta}_{i}) = 0. \notag
\end{equation}
The instrumental variable refinements, from now denoted as SRIVC iterations, are denoted using the iterate counter $s$, as
	\begin{equation}
	\begin{split}
		\theta_{i,s+1}^{l+1} &= \left[\sum_{k=1}^N \alpha_{N-k} \hat{\varphi}_{i,\textnormal{f}} (t_k, \theta_{i,s}^{l+1}) \varphi_{i,\textnormal{f}}^\top(t_k, \theta_{i,s}^{l+1})\right]^{-1} \\
        &\times\left[\sum_{k=1}^N \alpha_{N-k} \hat{\varphi}_{i,\textnormal{f}}(t_k, \theta_{i,s}^{l+1}) \tilde{y}_{i,\textnormal{f}}(t_k, \theta_{i,s}^{l+1})\right].
  		\label{eq:srivc}
	\end{split}
	\end{equation}
\ref{app:Deriv} shows how \eqref{eq:opt2} fits the pseudolinear regression framework with \eqref{eq:pseudolinear} and how this leads to SRIVC iterations in \eqref{eq:srivc}. In \eqref{eq:srivc}, the subscripts $i$ and $l$ refer to the current block-coordinate descent iteration. The filtered residual output, regressor vector and instrument vector, indicated with subscript $\mathrm{f}$, are respectively given by 
\begin{align}
			\label{eq:srivcoutput1}
		\tilde{y}_{i,\textnormal{f}}(t_k,\theta_{i,s}^{l+1})&= \frac{1}{A_i(p,\theta_{i,s}^{l+1})}\tilde{y}_i(t_k),  \\
		\begin{split} {\varphi}_{i,\textnormal{f}}(t_k,\theta_{i,s}^{l+1})&= \bigg[
			\dfrac{-p}{A_i(p,\theta_{i,s}^{l+1})}\tilde{y}_i(t_k) \:\:\:\: \dots \:\:\:\: \dfrac{-p^{n_i}}{A_i(p,\theta_{i,s}^{l+1})}\tilde{y}_i(t_k) \\
   	\label{eq:srivcregressor1}
&\quad \dfrac{1}{A_i(p,\theta_{i,s}^{l+1})} u(t_k) \:\:\:\: \dots \:\:\:\: \dfrac{p^{m_i}}{A_i(p,\theta_{i,s}^{l+1})} u(t_k)			
		\bigg]^\top,
		\end{split}  \\
  \begin{split}
  \hat{{\varphi}}_{i,\textnormal{f}} (t_k,\theta_{i,s}^{l+1} ) &= \bigg[
		\dfrac{-p}{A_i(p,\theta_{i,s}^{l+1})}\hat{x}_i(t_k) \:\:\:\: \dots \:\:\:\: \dfrac{-p^{n_i}}{A_i(p,\theta_{i,s}^{l+1})} \hat{x}_i(t_k) \\
  	\label{eq:srivcinstrument1}
&\quad \dfrac{1}{A_i(p,\theta_{i,s}^{l+1})}u(t_k) \:\:\:\: \dots \:\:\:\: \dfrac{p^{m_i}}{A_i(p,\theta_{i,s}^{l+1})} u(t_k)
		\bigg]^{\top},
	\end{split} 
 \end{align}
where $\hat{x}_i(t_k)$ is the estimate of $x_i(t_k)$. This estimate is based on the latest parameter estimate of the previous time-sample, denoted by $\bar{\theta}_{i} (t_{k-1})$, i.e.,
\begin{equation}
\hat{x}_i(t_k) = \frac{B_i(p, \bar{\theta}_{i} (t_{k-1}) )}{A_i(p, \bar{\theta}_{i} (t_{k-1}) )} u(t_k). \notag 
\end{equation}
To clarify, the refinement step refers to updating the denominator polynomial of the prefiltering step in \eqref{eq:srivcoutput1} to \eqref{eq:srivcinstrument1}. Hence, the filtered output, regressor vector, and instrument vector, are updated based on the estimates obtained from the previous iterate. The instrument vector is similar to the regressor vector but uses $\hat{x}_i$ instead of $\tilde{y}_i$, i.e., an estimate of the output of the submodel instead of using the noisy measured variable.

\begin{remark}
	The regressor \eqref{eq:srivcregressor1} and the instrument \eqref{eq:srivcinstrument1} require the computation of derivatives up to order $m_i$ for the input and $n_i$ for the residual output. Note that the order of the required derivatives is significantly lower compared to the case where a single unfactored transfer function would have been used. In this scenario, given that $\sum_{i=1}^{K}n_i = n$, obtaining the derivative of the $n$th order of the output is necessary for standard recursive continuous-time identification methods, along with a substantially higher derivative of the input. \label{rem:derivativeorder}
\end{remark}

\begin{remark}
	Note that the expressions in \eqref{eq:srivc} to \eqref{eq:srivcinstrument1} give the standard SRIVC iterations when $\tilde{y}(t_k) = y(t_k)$ (i.e., when $K=1$ and all the modes of the composite transfer function are estimated jointly) and $\alpha_{N-k}=1$ for all integer $k$ less than $N$ (i.e., when there is no forgetting factor).
\end{remark}
	
\begin{remark}
The expressions provided in \eqref{eq:srivcoutput1}, \eqref{eq:srivcregressor1} and \eqref{eq:srivcinstrument1} require prefiltering the data, relying implicitly on the commutativity property given by
	\begin{equation}
		\label{eq:commutativity}
		cG(p)x(t) = G(p)[cx(t)],
	\end{equation}
	where $c$ is a constant. It is not recommended to prefilter the data by transfer functions when the parameters exhibit rapid temporal variations, since the property \eqref{eq:commutativity} does not hold in general when $c$ is time-varying \citep{toth2010modeling}. However, it is important to note that \eqref{eq:commutativity} is a useful approximation in cases where the parameters exhibit slow temporal variations.
\end{remark}

Instead of computing \eqref{eq:srivc}, an equivalent recursion is derived which does not involve the summation operator over all available samples and allows to achieve an update of the estimate after every incoming time sample. This recursion involves computing	
	\begin{align}
 \begin{split}
		K_{i,s+1}^{l+1} (t_{N}) &= \frac{\bar{P}_{i}(t_{N-1}) \hat{\varphi}_{i,\textnormal{f}}(t_N,\theta_{i,s}^{l+1})}{ \lambda + \varphi^{\top}_{i,\textnormal{f}}(t_N,\theta_{i,s}^{l+1}) \bar{P}_{i}(t_{N-1})  \hat{\varphi}_{i,\textnormal{f}}(t_N,\theta_{i,s}^{l+1})}, 		\label{eq:RSRIVC_K} 
 \end{split} \\
 \begin{split}
		\theta_{i,s+1}^{l+1} (t_{N}) &= \bar{\theta}_{i} (t_{N-1}) + K_{i,s+1}^{l+1} (t_{N}) \\ &  \times \left(  \tilde{y}_{i,\textnormal{f}}(t_N,\theta_{i,s}^{l+1}) - {\varphi}_{i,\textnormal{f}}^\top(t_N,\theta_{i,s}^{l+1}) \bar{\theta}_{i} (t_{N-1}) \right),
		\label{eq:RSRIVC_theta}
   \end{split}
	\end{align}	
and
	\begin{align}
		P_{i,s+1}^{l+1}(t_{N}) &= \frac{1}{\lambda} \big( \bar{P}_{i}(t_{N-1}) - K_{i,s+1}^{l+1} (t_{N}) {\varphi}^{\top}_{i,\textnormal{f}}(t_{N},\theta_{i,s}^{l+1}) \bar{P}_{i}(t_{N-1})  \big),
		\label{eq:RSRIVC_P}
	\end{align}	
where $\bar{\theta}_{i} (t_{N-1})$, and $\bar{P}_{i}(t_{N-1})$ denote the last parameter estimate and inverse of the matrix being inverted in \eqref{eq:srivc} at the previous time instance $t_{N-1}$ after the $\mathcal{A}$-iterations over $l$ and $i$, and after the SRIVC iterations using $s$. The derivation of this recursion is presented in \ref{app:Deriv2}. Note that $\bar{P}_{i}(t_{N-1})$ is interpreted as an approximate covariance matrix of the resulting estimate, see \citet{young2015refined} and \citet{pan2020efficiency} for more details. A summary of the algorithm is provided in \Cref{subsec:summary}.
		

\subsection{Consistency analysis with $\alpha_{k} = 1$}
First, the conditions under which the additive RSRIVC approach produces a consistent estimator are analyzed. To this end, consider the following assumptions to facilitate the analysis, theorem, and related remarks. These assumptions are standard in the SRIVC analysis literature, see, e.g., \cite{pan2020consistency,gonzalez2024statistical,gonzalez2023closedloop}. The scenario with $\alpha_{k} = 1$ allows to employ a similar reasoning as the non-recursive SRIVC approach as presented in \cite{pan2020consistency}, although special caution is taken to address the time-varying nature of the models used for constructing the instrument and regressor vectors.
\begin{assumption}[BIBO stability]
	\label{ass:assumption1}
      The true system and the estimated models are uniformly bounded-input, bounded-output (BIBO) stable, i.e., for any initial time instant $t_{0}$ and bounded input signal, the output sequence $\{y(t_k)\}_{k\geq 0}$ is bounded \citep{rugh1996linear}.
 \end{assumption}
 \begin{assumption}[Persistence of excitation]
	\label{ass:assumption2}
	The input sequence is persistently exciting of order no less than $2n$.
\end{assumption}
\begin{assumption}[Coprimeness and sampling frequency]
	\label{ass:assumption3}
    The additive subsystems pertaining to the true system and the models obtained through the descent iterations do not have zero-pole cancellations. Furthermore, the sampling frequency is more than twice the largest positive imaginary part of the zeros of $\prod_{i=1}^K A_i(p,\theta_{i,s}^l)A_i^*(p)$. The subscript $s$ and superscript $l$ relate to the SRIVC iterations and block-coordinate descent iterations, respectively, defined in \Cref{sec:recursive}.
\end{assumption}
\begin{assumption}[Independence and stationarity]
	\label{ass:assumption4}
	The input sequence $\{u(t_k)\}$ is quasi-stationary \citep[p. 34]{ljung1998system}, the disturbance $\{v(t_l)\}$ is stationary, and they are mutually independent for all integers $k$ and $l$.
\end{assumption}
\begin{assumption}[Parameter trajectories]
    \label{ass:assumption5}
    The changes in the parameters occur only at the sampling instants $t_k$, i.e., the parameter trajectories are piecewise constant.
\end{assumption}
\begin{theorem} \label{thm:Consistency_theorem}
    Consider the proposed estimator for the open-loop setting in \Cref{fig:ol_setup} with $\alpha_{k}=1$ for all $k$, and suppose the following assumptions, Assumptions \ref{ass:assumption1} to \ref{ass:assumption5} hold. In addition, assume that the SRIVC iterations of the estimator for all $N$ sufficiently large converge, where the converged parameter is denoted by $\theta_{i,\infty}^{l+1}$. Then, if both the proposed model and system converge as $N\to \infty$ to linear and time-invariant transfer functions, then their limits are equal with probability 1 and generically consistent\footnote{A statement $s$, which depends on the elements $\rho$ of some open set $\Omega \subseteq \mathbb{R}^n$, is \textit{generically true with respect to }$\Omega$ \citep{soderstrom1983instrumental} if the set $\{\rho\in \Omega| s(\rho) \textnormal{ is not true}\}$ has Lebesgue measure zero in $\Omega$.} with respect to the limiting system and model parameters.
\end{theorem}
The proof is provided in \ref{Appendix:Deriv3}.

\begin{remark}
Note that consistency is only considered for $\alpha_{k}=1$ for all $k$. This is a special case of \eqref{eq:FF}. In case \eqref{eq:FF} is used with a different constant-rate-forgetting factor $\lambda < 1$, the effect of noise does not disappear for $N \rightarrow \infty$. Therefore, considering more data does not reduce the influence of measurement noise making the estimator not consistent. Variable-rate-forgetting factors may be considered, ensuring consistent estimators (see \citet{bruceConvergenceConsistencyRecursive2020} for details); however, these considerations are beyond the scope of this work.
\end{remark}

\subsection{Summary iterative recursive estimation algorithm} \label{subsec:summary}
The proposed additive RSRIVC algorithm involves nested iterations in $l$, $i$, and $s$. To facilitate implementation, the nested iterations are discussed in this subsection.

First of all, the recursive nature allows to update the estimate after every incoming sample. Therefore, these iterations will be performed at the discrete times $t_k$, where $k$ denotes the count in the discrete-time domain. At each of these points in time, the $\mathcal{A}$-iterations are performed $M_{l}$ times, see \Cref{subsec:Block_coordinate_descent}, i.e., the block-coordinate method runs through each of $K$ additive models for $M_l$ times. The current additive model is denoted by $i$ and the $\mathcal{A}$-iteration is denoted by $l$. Finally, within each of these parameter updates, $M_s$ SRIVC iterations are performed, with the current SRIVC iterate being indicated by $s$.

\begin{figure}[tb]
      \centering
      \includegraphics[trim={0mm 0mm 0mm 0mm},clip]{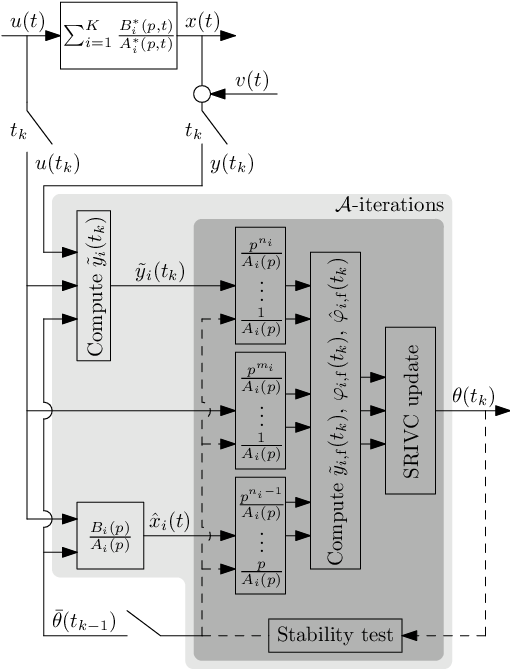}
	  \caption{Block diagram of the block-coordinate descent RSRIVC for open-loop systems, summarized by \Cref{alg:algorithm1}. The section featuring SRIVC iterations is highlighted \tikzsquare{MatlabGray70}, see \eqref{eq:srivc} to \eqref{eq:srivcinstrument1}, alongside the block containing the $\mathcal{A}$-iterations \tikzsquare{MatlabGray90}, described in \Cref{subsec:Block_coordinate_descent}.}
      \label{fig:RSRIVC_block_diagram}
   \end{figure}

To summarize, compute the recursive formulas in \eqref{eq:RSRIVC_K}, \eqref{eq:RSRIVC_theta} and \eqref{eq:RSRIVC_P} at each time step and for every additive model $G_i$ to solve the additive model identification problem in \eqref{eq:opt2}. This way, the proposed method solves \eqref{eq:opt2} recursively. The algorithm is presented in Algorithm \ref{alg:algorithm1}. Note that the regressor vector has an explicit dependence on the parameter vector of the other models being estimated; this dependence is implicit in the equations via the expression for the residual $\tilde{y}_{i}$. A block-diagram of the block-coordinate descent RSRIVC algorithm for additive systems is shown in \Cref{fig:RSRIVC_block_diagram}. Two distinct blocks are highlighted in this diagram, which refer to the SRIVC and the block-coordinate descent step. These are executed after every incoming sample. During these iterations, the quantities $\tilde{y}_{i,f}$, $\varphi_{i,f}$, $\hat{\varphi}_{i,f}$ are updated, which only involves filtering between $t_{k-1}$ and $t_{k}$. Optionally, a stability test is included which is further examined in Section \ref{sec:implementation}.

		\begin{algorithm}[tb]
		\caption{Recursive estimator for open-loop continuous-time systems in additive form}
		\begin{algorithmic}[1]
			\State 	\textbf{Initialization}: Set $M_{l}$, $M_{s}$, $\lambda$. Initialize $\bar{\theta}_{i} (t_{0})$, and $\bar{P}_i (t_{0})$ for each $i = 1, \ldots, K$, by using the first part of the input-output data $\{u(t_k),y(t_k)\}_{k=-N_{\mathrm{init}}}^{0}$ with any linear and time-invariant continuous-time system identification method. With this $\bar{\theta}_{i} (t_{0})$, compute ${\varphi}_{i,\textnormal{f}}$, $\hat{{\varphi}}_{i,\textnormal{f}}$, and $\tilde{y}_{i,\textnormal{f}}$ at $t_{0}$ by filtering the input-output data $\{u(t_k),y(t_k)\}_{k=-N_{\mathrm{init}}}^{0}$ through (26)-(28) using the denominator $A_{i}(p,\bar{\theta}_{i} (t_{0}))$.
			\For{$k = 1, \dots$}
 \State  $\theta_{i}^{1} (t_k) \gets \bar{\theta}_{i} (t_{k-1})$ for all $i$
\For{$l = 1, \dots, M_{l}$}
				\For{$i = 1, \dots, K$}
 \State  $\theta_{i,0}^{l+1} (t_k) \gets \theta_{i}^{l} (t_k)$
		\For{$s = 0, 1,\dots, M_{s}$}
			
			\State Parameter $\theta_{i,s+1}^{l+1} (t_k)$ update:
		\begin{align}
			K_{i,s+1}^{l+1} (t_k) &\gets \frac{\bar{P}_{i}(t_{k-1}) \hat{\varphi}_{i,\textnormal{f}}(t_k,\theta_{i,s}^{l+1})}{\lambda + \varphi^{\top}_{i,\textnormal{f}}(t_k,\theta_{i,s}^{l+1}) \bar{P}_{i}(t_{k-1})  \hat{\varphi}_{i,\textnormal{f}}(t_k,\theta_{i,s}^{l+1})}, \notag \\
			\theta_{i,s+1}^{l+1} (t_k) &\gets \bar{\theta}_{i} (t_{k-1}) + K_{i,s+1}^{l+1} (t_{k}) \notag  \\ &\times \left(  \tilde{y}_{i,\textnormal{f}}(t_k,\theta_{i,s}^{l+1}) - {\varphi}_{i,\textnormal{f}}(t_k,\theta_{i,s}^{l+1}) \bar{\theta}_{i} (t_{k-1}) \right), \notag \\
			P_{i,s+1}^{l+1} (t_k) &\gets \frac{1}{\lambda} \bar{P}_{i}(t_{k-1}) - \frac{1}{\lambda} K_{i,s+1}^{l+1} (t_{k}){\varphi}^{\top}_{i,\textnormal{f}}(t_k,\theta_{i,s}^{l+1}) \bar{P}_{i}(t_{k-1}). \notag 
		\end{align}
						
		\EndFor
        \State $\theta_{i}^{l+1} (t_k) \gets \theta_{i,M_s+1}^{l+1} (t_k)$
        \State $P_{i}^{l+1} (t_k) \gets P_{i,M_s+1}^{l+1} (t_k)$
		\EndFor		
		\EndFor
        \State  $\bar{\theta}_{i} (t_{k}) = \theta_{i}^{M_l+1} (t_k)$ for all $i$
        \State  $\bar{P}_{i} (t_{k}) = P_{i}^{M_l+1} (t_k)$ for all $i$
		\EndFor
			\State \textbf{Output}: the parameter estimates $\{ \bar{\theta}_i(t_k) \}_{k \geq 1}$.
		\end{algorithmic}
		\label{alg:algorithm1}
	\end{algorithm}
	
\begin{remark}
Note that the SRIVC refinements in $s$ are optional. In case the refinements are omitted, the optimization problem \eqref{eq:opt2} is not being solved until convergence at every time-step, but rather a quasi-Newton descent algorithm of the form \eqref{eq:srivc} is being implemented \citep{young2015refined}. Despite this, the iterations stem from the fact that solving \eqref{eq:opt2} and iterating over the additive models is a viable way to derive a model that fully minimizes the cost in \eqref{eq:op}.
\label{rem:simpler_RSRIVC}
\end{remark}	

\begin{remark}
 While not inherently restrictive to the method, it is practically advantageous if parameters change gradually compared to the sampling rate. In general, slower parameter variation improves the effectiveness of the methods as it allows reducing the number of iterations $M_{l}$ and $M_{s}$, which in  turn reduces computational burden and improves parameter adaptation efficiency and robustness.
    \label{rem:slowly_varying_params}
\end{remark}

\subsection{Closed-loop case}
The identical recursion as in \eqref{eq:RSRIVC_K}, \eqref{eq:RSRIVC_theta} and \eqref{eq:RSRIVC_P} applies to the closed-loop scenario, albeit with a different instrument vector. Employing an output error cost for system identification is known to introduce asymptotic bias in the presence of output measurement noise \citep{van1998closed}. Consequently, instrumental variable methods are used in the closed-loop setting to mitigate this bias \citep{gilson2011optimal,gonzalez2023closedloop}. Instead of \eqref{eq:opt2}, the instrumental variable solution is employed which is given by 
\begin{equation}
\begin{split}
    \theta_i^{l+1} \in \underset{\theta_i \in \Omega_i}{\textrm{sol}} \Bigg\{ 
&\sum_{k=1}^N \alpha_{N-k} \hat{\varphi}_{i,\textnormal{f}} (t_k, \theta_{i}) \\ &\times \left( \tilde{y}_{i,\textnormal{f}}(t_k, \theta_{i}) - \varphi_{i,\textnormal{f}}^{\top} (t_k, \theta_{i}) \theta_{i} (t_k) \right) = 0 \Bigg\}, \notag
\end{split}
\end{equation} 
 with $\hat{\varphi}_{i,\textnormal{f}}$ being the filtered instrument vector related to additive model $i$. Users have the flexibility to choose from various instrument vectors \citep{soderstrom1983instrumental}. In this context, the instrument is constructed as a noiseless version of the regressor, a choice recognized in similar identification contexts for yielding minimal asymptotic covariance among all instrumental variable estimators \citep{gilson2008instrumental,boeren2018optimal,mooren2023online,gonzalez2023identification}. To this end, the instrument \eqref{eq:srivcinstrument1} is rewritten as
 	\begin{align}
  \begin{split}
		\hat{{\varphi}}_{i,\textnormal{f}} (t_k,\theta_{i,s}^{l+1} ) &= \bigg[
		\dfrac{-p}{A_i(p,\theta_{i,s}^{l+1})} \frac{B_i(p, \bar{\theta}_{i} (t_{k-1}) )}{A_i(p, \bar{\theta}_{i} (t_{k-1}) )} \:\:\:\: \dots \\ & \quad \dfrac{-p^{n_i}}{A_i(p,\theta_{i,s}^{l+1})} \frac{B_i(p, \bar{\theta}_{i} (t_{k-1}) )}{A_i(p, \bar{\theta}_{i} (t_{k-1}) )} \\
& \quad \dfrac{1}{A_i(p,\theta_{i,s}^{l+1})} \:\:\:\: \dots \:\:\:\: \dfrac{p^{m_i}}{A_i(p,\theta_{i,s}^{l+1})} 
		\bigg]^{\top} z(t_{k}),
  	\label{eq:srivcinstrument1cl}
  \end{split}
	\end{align}
however, instead of $z(t_{k}) = u(t_k)$, the noiseless input is estimated as $z(t_{k}) = S_{uo}(q,t_k) r(t_k)$, where the sensitivity function $S_{uo}(q,t_{k}) = C_{\textnormal{d}}(q)/[1 + C_{\textnormal{d}}(q) G_{\textnormal{d}}(q,t_k)]^{-1}$. Hence, the instrument now depends on an estimate of the control input, computed by means of a model of the sensitivity and the known setpoint $r(t_k)$.

\subsection{Marginally stable systems}
\label{subsec:marginally}
Mechatronic position systems, frequently encountered in engineering applications, lend themselves to a decomposition comprising both rigid body and flexible modes. Notably, the rigid body modes are considered as double integrators, leading to eigenvalues precisely situated at the origin. The denominator polynomial of such marginally stable system cannot be represented in an anti-monic form as in \eqref{eq:apt2} and~\eqref{eq:bpt2}. To this end, the proposed parametrization is slightly adapted to accommodate for the estimation of marginally stable systems as follows.

Consider the system \eqref{eq:systemadditive}, but where the first submodel has $\ell$ poles at the origin. In other words, let $G^*_1(p,t) = B_1^*(p,t)/[p^\ell A_1^*(p,t)]$, where $A_{1}^*(p,t)$ and $B_1^*(p,t)$ are co-prime, and have the same form as in \eqref{eq:apt} and \eqref{eq:bpt}. Let the remaining submodels be parameterized as originally defined in \Cref{subsec:problem_formulation}. For either open or closed-loop variants of the proposed approach, the computation of the gradient of each submodel with respect to their respective parameter vector is required. For $G_1(p)$, this computation results in the modified instrument vector
\begin{equation}
\label{eq:instrument_int}
    \begin{split}
	\hat{{\varphi}}_{1,\textnormal{f}} (t_k,\theta_{1,s}^{l+1} )& = \bigg[\frac{-p }{p^{\ell} A_{1}(p,\theta_{1,s}^{l+1}) } \frac{B_1(p, \bar{\theta}_{i} (t_{k-1}) )}{A_1(p, \bar{\theta}_{i} (t_{k-1}) )}
\:\:\:\: \dots \\ &\quad \frac{-p^{n_1}}{p^{\ell} A_{1}(p,\theta_{1,s}^{l+1})} \frac{B_1(p, \bar{\theta}_{i} (t_{k-1}) )}{A_1(p, \bar{\theta}_{i} (t_{k-1}) )}  \\ &\quad \frac{1}{p^{\ell}A_{1}(p,\theta_{1,s}^{l+1})} \:\:\:\: \dots \:\:\:\: \frac{p^{m_1}}{p^{\ell}A_{1}(p,\theta_{1,s}^{l+1})}\bigg]^\top z(t_k),        
    \end{split}
\end{equation}
where $z(t_k)=u(t_k)$ for the open-loop algorithm, and $z(t_k)=S_{uo}(q,t_k) r(t_k)$ for the closed-loop variant. On the other hand, the model residual retains the same form as in \eqref{eq:ytilde}, with the filtered residual output given by \eqref{eq:pseudolinear}, but with the regressor vector now expressed as
\begin{align}
\begin{split}
   {\varphi}_{1,\textnormal{f}} (t_k,&\theta_{1,s}^{l+1} )  = \bigg[\frac{-p}{A_{1}(p,\theta_{1,s}^{l+1})} \tilde{y}_1(t_k)  \:\:\:\: \dots \frac{-p^{n_1}}{A_{1}(p,\theta_{1,s}^{l+1})} \tilde{y}_1(t_k) \\ & 
	\frac{1}{p^\ell A_{1}(p,\theta_{1,s}^{l+1})}u(t_k) \:\:\:\: \dots  \:\:\:\: \frac{p^{m_1}}{p^\ell A_{1}(p,\theta_{1,s}^{l+1})} u(t_k) \bigg]^{\top}.
\label{eq:regressor_int}
\end{split}
\end{align}
By computing the iterations in \eqref{eq:srivc} with the first instrument and regressor vectors given by \eqref{eq:instrument_int} and \eqref{eq:regressor_int} respectively, a direct extension of the proposed method for identifying marginally stable systems in additive form is obtained. Aside from these modifications to the first submodel, the algorithm remains unchanged.
\begin{remark}
Assumption \ref{ass:assumption1} is exclusively employed in the proof for generic consistency with uniformly stable systems. This excludes marginally stable systems, yet it does not constrain the proposed approach.
\end{remark}

\vspace{3mm}
\subsection{Practical aspects for implementation}
\vspace{2mm}
\label{sec:implementation}
Next, several practical aspects of the additive RSRIVC implementation are investigated. First, the initialization process of the algorithm is discussed, followed by a discussion of the robustness of the algorithm. Subsequently, to alleviate the computational load associated with the approach, output subsampling is examined.


A proper initialization of the algorithm is important, particularly the initial estimate $\bar{\theta}_{i}(t_{0})$, which should be selected carefully, as it influences the initial values of ${\varphi}_{i,\textnormal{f}}$, $\hat{{\varphi}}_{i,\textnormal{f}}$, and $\tilde{y}_{i,\textnormal{f}}$ at $t_{0}$. Inadequate initialization can lead to divergence. The parameters $\bar{\theta}_{i} (t_{0})$ and covariance $\bar{P}_i (t_{0})$ can be obtained through any offline, non-recursive method using the input-output data $\{u(t_k),y(t_k)\}_{k=-N_{\mathrm{init}}}^{0}$. For example, approaches such as those described in \cite{gonzalez2023parsimonious} or \cite{gonzalez2023identification} may be utilized.

In the offline SRIVC method, ensuring model stability is vital, especially in challenging conditions like low Signal-to-Noise Ratio (SNR) or mismatched model orders \citep{ha2014ensuring}. Typically, stability is monitored and enforced throughout iterations \citep{garnier2015direct}. In RSRIVC, where both prefilter and auxiliary model depend on preceding estimates, stability verification is more intricate. For slowly-varying systems, checking characteristic equation poles and reflecting right-half plane poles to the left suffices for confirming bounded-input, bounded-output stability under specific conditions \citep{padilla2019identification}. Notably, in time-varying parameter models, uniform exponential stability is assured by assuming slow parameter changes and keeping system matrix eigenvalues on the left half plane. RSRIVC incorporates a projection algorithm to guarantee the stability of the prefilter and auxiliary model, reflecting unstable poles into the left half plane while preserving the magnitude characteristics. This approach addresses potential stability issues that may arise, particularly at the beginning of the data record.

The algorithm presented is theoretically sound, however, the recursion in \eqref{eq:RSRIVC_P} may not be numerically reliable \citep[Chapter 11]{ljung1983theory}. Since $P(t_k)$ is essentially computed by successive subtractions, round-off errors can accumulate and make $P(t_k)$ ill-conditioned. Other continuous-time recursive methods based on instrumental variables \citep{padilla2017recursive,pan2021identification} incorporate a covariance matrix $Q(t_k)$ that is added to $P(t_k)$ at each iteration. The proposed algorithm does not include this matrix since it is not compatible with the coordinate descent interpretation of the iterations in \Cref{alg:algorithm1}. However, from a practical perspective, including this matrix can help with the convergence of the parameters. To achieve smoother estimates, a constant prefilter can replace the adaptive filter in the proposed approach, though this may prevent SRIVC iterations from converging to a local minimum, affecting the global convergence of the block coordinate descent.

Subsampling data may be valuable to managing computational complexity without compromising significant accuracy. This facilitates the implementation of recursive algorithms in resource-constrained environments. Subsampling should be performed cautiously to ensure that the subsampled data still captures the smallest timescale for a reliable estimate. Therefore, it is primarily applicable when the timescale of the system dynamics is considerably shorter than the sampling time. When subsampling the output, it is advised to take all input samples into account for the estimation of the filtered input and filtered auxiliary variables. Hence, the block-coordinate descent algorithm and additive RSRIVC iterations may be performed at this lower sampling rate. Furthermore, as hinted in \Cref{rem:simpler_RSRIVC}, it is not always necessary to perform multiple iterations. In case the parameters of the system vary slowly with respect to the sampling period, typically no refinements are necessary and a single $\mathcal{A}$-iteration can achieve accurate estimates.

\vspace{2mm}
\section{Numerical Simulation}
\vspace{2mm}
\label{sec:simulations}
In this section, the proposed method is illustrated using the following additive system
\begin{equation}
    G^*(p,t)=\sum_{i=1}^3 \frac{b_{i,0}^*(t)}{a_{i,2}^*(t) p^2 +a_{i,1}^*(t) p +1}, \notag 
\end{equation}
with initial DC-gains $\begin{bmatrix}
    b_{1,0}^*(0) & b_{2,0}^*(0) & b_{3,0}^*(0)
\end{bmatrix} = \begin{bmatrix}
    3 & 0.5 & 1
\end{bmatrix}$, and associated denominator parameters $\Big[
    a_{1,2}^*(0) $ $ a_{2,2}^*(0) \:\:\:\: a_{3,2}^*(0)
\Big]$ $= \begin{bmatrix}
    0.25 & 0.04 & 0.0025
\end{bmatrix}$, and $\Big[
    a_{1,1}^*(0) \:\:\:\: a_{2,1}^*(0) \:\:\:\: a_{3,1}^*(0)
\Big]$ $=$ $\begin{bmatrix}
    0.25 & 0.01 & 0.01
\end{bmatrix}$. The output $y(t_k)$ is measured at a sampling frequency of $20$ [Hz] and is contaminated by a zero-mean Gaussian white noise with a variance of 0.01. The system is operated in closed-loop with feedback controller
\begin{equation}
K(q) = \frac{0.02329 q^2 - 0.2058 q + 0.00454}{q^2 - q}, \notag     
\end{equation}
   \begin{figure}[t]
      \centering
      \includegraphics[trim={0mm 0mm 0mm 0mm},clip,width=.88\columnwidth]{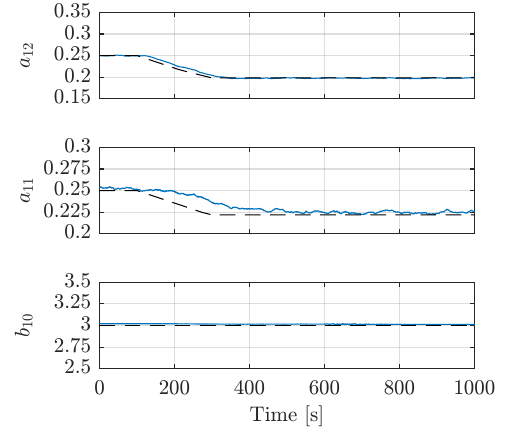}
	  \caption{Parameter estimates \tikzline{MatlabBlue} and true parameters \tikzdashedline{black} related to the first submodel with decreasing denominator parameters.}
      \label{fig:sim_param1}
   \end{figure}

   \begin{figure}[t]
      \centering
      \includegraphics[trim={0mm 0mm 0mm 0mm},clip,width=.88\columnwidth]{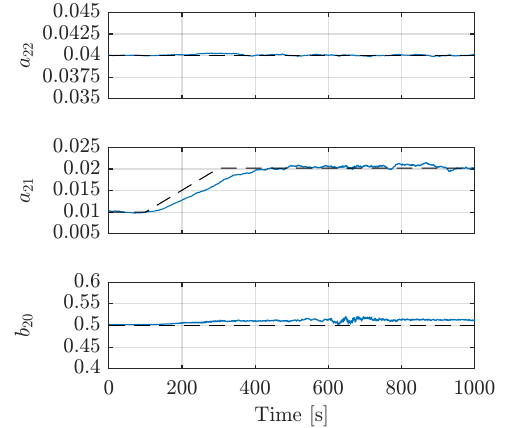}
    \caption{Parameter estimates \tikzline{MatlabBlue} and true parameters \tikzdashedline{black} related to the second submodel with decreasing damping coefficient $\zeta_2$, which corresponds to an increasing $a_{21}$.}
    \label{fig:sim_param2}
   \end{figure}

   \begin{figure}[t]
      \centering
      \includegraphics[trim={0mm 0mm 0mm 0mm},clip,width=.88\columnwidth]{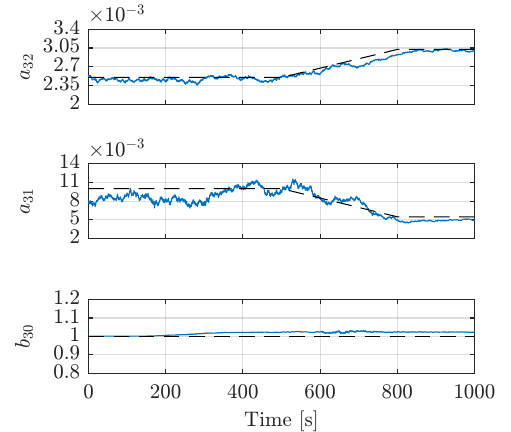}
    \caption{Parameter estimates \tikzline{MatlabBlue} and true parameters \tikzdashedline{black} related to the third submodel with decreasing natural frequency $\omega_3$, which corresponds to an increasing $a_{32}$ and decreasing $a_{31}$.}
      \label{fig:sim_param3}
   \end{figure}
\noindent and the setpoint equals $r(t) =  3 \sin{(0.005 t)}+\sin{(2 t)}+\sin{(5 t)}+\sin{(17.5 t)}$. The algorithm is initialized with covariance matrices $\bar{P}_i(t_{N_{1}}) = \mathrm{diag}(10^{-3},10^{-3},10^{-5})$, and forgetting factor $\lambda = 0.999$. The estimator is initialized at a random system with parameters $\bar{\theta}_i(t_{N_{1}}) = \theta_i^{*}(0) (1+\mathcal{U}(-0.02,0.02))$, where $\mathcal{U}(a,b)$ denotes a uniform distribution with lower limit $a$ and upper limit $b$.

First, in the interval $t \in [100, 300]$ [s], the parameters $a_{1,1}$, $a_{1,2}$, and $a_{2,1}$, are varied in such a way that the corresponding natural frequency $\omega_{1}$ and damping coefficient $\zeta_{2}$ decrease, c.f., \eqref{eq:mechsys}. Subsequently, in the interval $t \in [500, 800]$ [s], the denominator parameters of the third model, i.e., $a_{3,2}$ and $a_{3,1}$ are varied.

\begin{figure}
      \centering
      \includegraphics[trim={0mm 0mm 0mm 0mm},clip,width=.88\columnwidth]{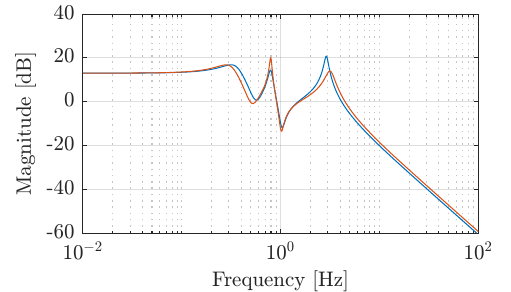}
	  \caption{Plant estimate at $t = 50$ [s] \tikzline{MatlabBlue} and at  $t = 950$ [s] \tikzline{MatlabRed}. Clearly, all resonances changed. Note that the natural frequency of the first mode decreased and the damping of the second mode decreased. A combination of the two occurred at the third mode.}
      \label{fig:sim_models}
   \end{figure}

\Cref{alg:algorithm1} is deployed to estimate the time-varying parameters, where no SRIVC iterations are performed and only a single coordinate descent iteration is performed at every incoming sample, i.e., the two loops in the algorithm are omitted as $s=0$ and $M_l = 1$. Since the system is operating in closed loop, the instrument is chosen as \eqref{eq:srivcinstrument1cl}. The obtained estimates and the true parameters are depicted for each of the submodels in Figures \ref{fig:sim_param1} to \ref{fig:sim_param3}. The estimated model at $t=50$ [s] and $t=950$ [s] are depicted in \Cref{fig:sim_models}. From the figures is concluded that no further refinements and gradient descent steps are required to achieve an accurate tracking of the parameters.

\section{Experimental validation on an overactuated and oversensed flexible beam setup}
\label{sec:experiment}
\begin{figure}[t]
	\centering{
	\setlength{\fboxsep}{-1pt}%
\setlength{\fboxrule}{1pt}%
    \fbox{\includegraphics[width=243pt]{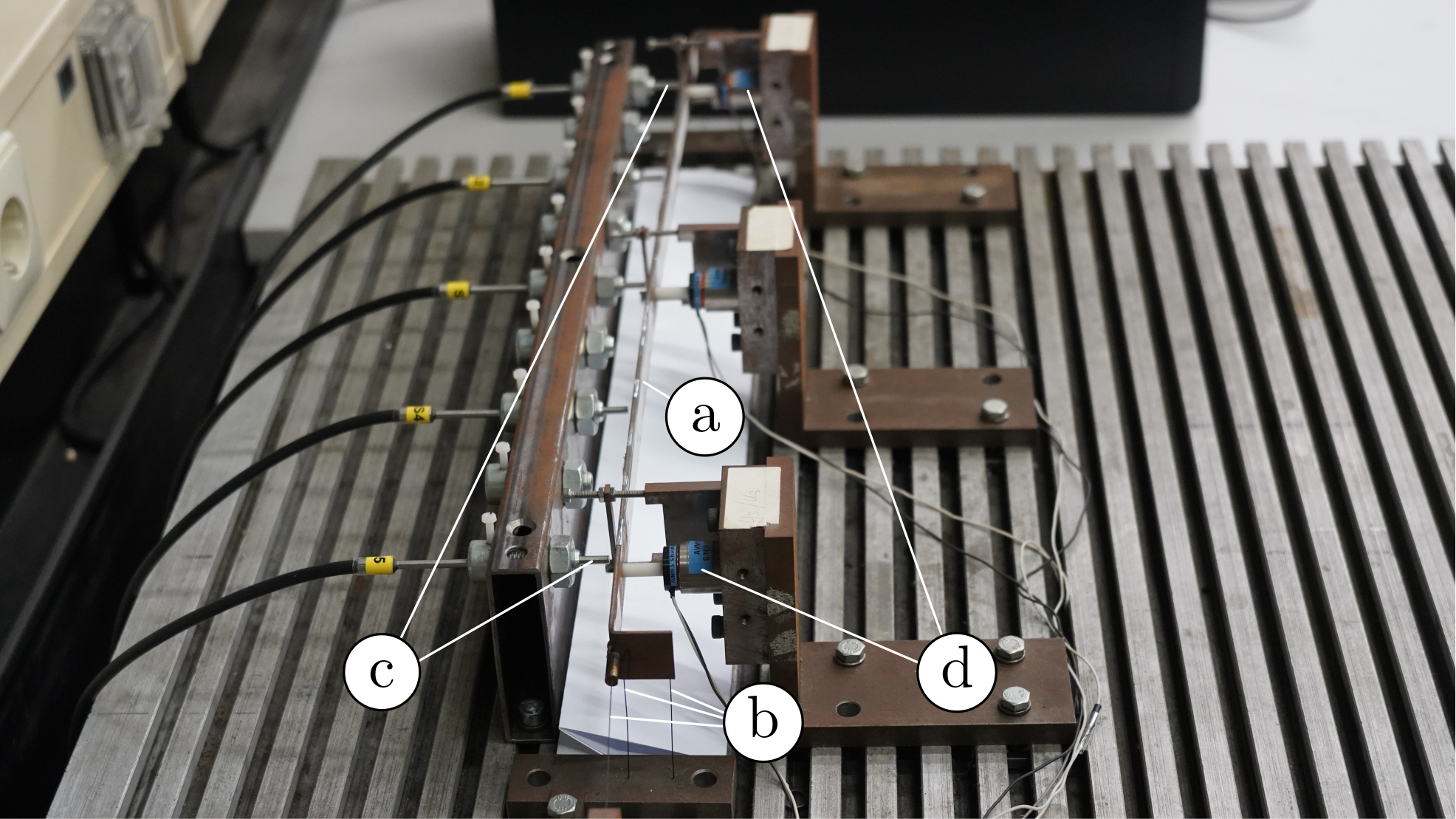}}
\caption{Prototype experimental flexible beam setup. The moving part is indicated by \encircle{a} and is suspended by wire flexures \encircle{b}. The deflection is measured with five contactless fiber optic sensors, of which two are used \encircle{c} and the setup is actuated with three current-driven voice coils of which the outer two are used \encircle{d}.}
		\label{fig:beam1}
    }
\end{figure}

In this section, the method is tested on an experimental setup. To this end, consider the setup shown in \Cref{fig:beam1}. The system consists of a thin flexible steel beam of $500 \times 20 \times 2$ [mm]. It is equipped with five contactless fiberoptic sensors and three voice-coil actuators and is suspended by wire flexures, leaving one rotational and one translational direction unconstrained. The system is operating at a sampling frequency of $4096$ [Hz]. The second and fourth sensor are not used for the conducted experiments. 

\begin{figure}[t]
	\centering{
		\includegraphics[width=0.48\textwidth]{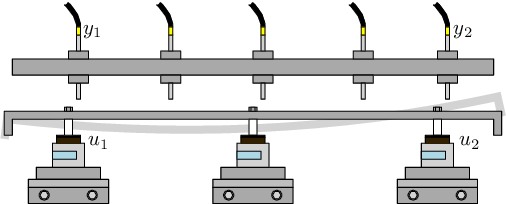}
		\caption{Schematic illustration of the experimental setup. The three actuators are visible on the bottom part of the image and the five sensors at the top. During operation, the beam translates, rotates and exhibits internal flexible behavior. The inputs and outputs are transformed as $u_{1} = u_{2} = \tfrac{1}{2} u$ and $y = \tfrac{1}{2} \left( y_{1} + y_{2} \right)$, respectively.}
		\label{fig:beam2}
    }
\end{figure} 

The main aim is to estimate the SISO system $G^{*}$ between the input $u$, exciting the system at the outer ends of the beam equally, see \Cref{fig:beam1,fig:beam2}, and the output $y$ which is the average deflection of the outer ends. This average deflection is controlled by a feedback controller and the setpoint is a square wave with a frequency of $2$ [Hz]. By means of a second internal control loop, the stiffness of the beam is artificially manipulated such that the observed behavior is time-varying, see \citet{classens2021fault,classens2022fault} for details. 

\begin{figure}
      \centering
      \includegraphics[trim={0mm 0mm 0mm 0mm},clip,width=.88\columnwidth]{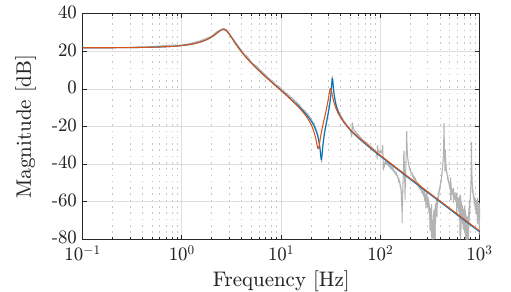}
	  \caption{Frequency response function of the nonmanipulated system \tikzline{MatlabGray70}, the plant estimate at $t = t_{N_{1}} = 0$ [s] \tikzline{MatlabBlue}, and at the plant estimate at $t = 135$ [s] \tikzline{MatlabRed}.}
      \label{fig:exp_bode}
   \end{figure}

Lightly-damped systems such as the one under study are well-suited to be described in a modal representation, see Example \ref{example:mechsys}, i.e., 
\begin{equation}
    G^*(p)=\sum_{i=1}^K \frac{b_{i,0}^*}{p^2/\omega_i^2 + 2(\zeta_i/\omega_i) p + 1}, \notag 
\end{equation}
where $\zeta_i$ and $\omega_i$ represent the damping coefficients and natural frequencies of the flexible modes respectively, see \citet{gawronski2004advanced}. The flexible beam fits the modal description, and the system is approximated with $K=2$ modes. The first mode is the suspension mode of the system and the second mode is the first internal flexible mode which is artificially manipulated. Figure \ref{fig:exp_bode} shows the model of the system in the frequency domain, including the neglected higher order modes, and the model at $t_{0}$ after initializing the system. Two distinct forgetting factors are used tailored to the distinct modes, namely $\lambda_1 = 0.9999$ and $\lambda_2 = 0.999$, and the approximate covariance matrices are initialized at $\bar{P}_{1}(t_{0}) = \bar{P}_{2}(t_{0}) = \mathrm{diag}(10^{-11}, 10^{-6}, 10^{-1})$.

\begin{figure}
      \centering
      \includegraphics[trim={0mm 0mm 0mm 0mm},clip,width=.88\columnwidth]{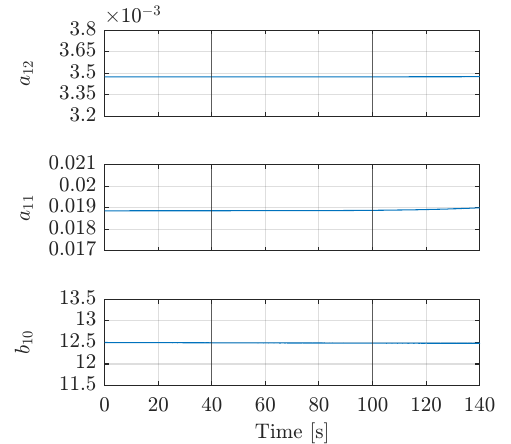}
	  \caption{Parameter estimate related to the first submodel. This first mode relates to the suspension of the flexible beam setup which remains unaltered throughout the experiment explaining the graphs.}
      \label{fig:exp_param1}
   \end{figure}

\begin{figure}
      \centering
      \includegraphics[trim={0mm 0mm 0mm 0mm},clip,width=.88\columnwidth]{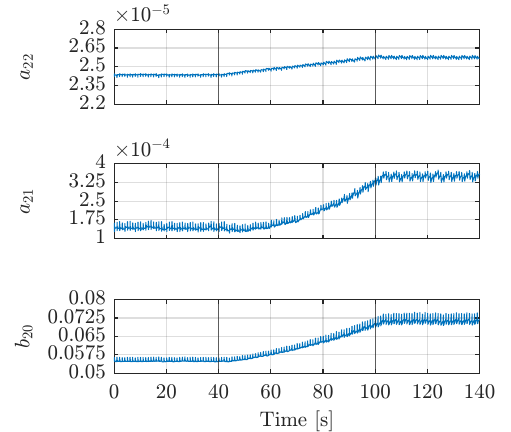}
	  \caption{Parameter estimate related to the second submodel. This mode is artificially manipulated between $t=40$ [s] and $t=100$ [s], which explains the changing parameters.}
      \label{fig:exp_param2}
   \end{figure}

An experiment is conducted where the internal flexible mode is actively manipulated from $t \in [40, 100]$ [s]. \Cref{alg:algorithm1} is deployed to estimate the time-varying behavior, where no SRIVC iterations are performed and only a single coordinate descent iteration is performed at every incoming sample, i.e., $s=0$ and $M_l=1$. Since the system is operating in closed loop, the instrument is chosen as \eqref{eq:srivcinstrument1cl}. The obtained estimates are depicted for each of the two submodels in \Cref{fig:exp_param1,fig:exp_param2}, and the estimated model at $t=0$ [s] and $t=135$ [s] is depicted in \Cref{fig:exp_bode}. From the figures is concluded that the shifting resonance mode is effectively observed, and the first mode does not exhibit parameter variations. The small fluctuations in the estimate of the second mode are at the frequency of the setpoint.

\section{Conclusions}
\label{sec:conclusions}
This article adresses a recursive method to identify additive systems for both open and closed-loop setups. The proposed algorithms, based on a block-coordinate descent with refined instrumental variables, are capable to track time-varying parameters of continuous-time systems in real-time. This allows tracking of more parsimonious and physically-relevant model representations, particularly for mechanical systems. Practical aspects are provided and finally the effectiveness of the proposed approach is illustrated through numerical validation and application on a real-time setup.

\section*{CRediT authorship contribution statement}
\textbf{Koen Classens:} Conceptualization, Methodology, Software, Formal Analysis, Validation, Writing - Original Draft, Writing - Review \& Editing, \textbf{Rodrigo A. Gonz\'alez:} Conceptualization, Methodology, Formal Analysis, Writing - Original Draft, Writing - Review \& Editing \textbf{Tom Oomen:} Writing - Review \& Editing, Supervision, Funding acquisition

\section*{Declaration of competing interest}
The authors declare that they have no known competing financial interests or personal relationships that could have appeared to influence the work reported in this paper.

\section*{Acknowledgement}
This work is supported by ASML Research, Veldhoven, the Netherlands.

   \bibliographystyle{elsarticle-harv} 
    \bibliography{references}

\begin{thebibliography}{44}
\expandafter\ifx\csname natexlab\endcsname\relax\def\natexlab#1{#1}\fi
\providecommand{\url}[1]{\texttt{#1}}
\providecommand{\href}[2]{#2}
\providecommand{\path}[1]{#1}
\providecommand{\DOIprefix}{doi:}
\providecommand{\ArXivprefix}{arXiv:}
\providecommand{\URLprefix}{URL: }
\providecommand{\Pubmedprefix}{pmid:}
\providecommand{\doi}[1]{\href{http://dx.doi.org/#1}{\path{#1}}}
\providecommand{\Pubmed}[1]{\href{pmid:#1}{\path{#1}}}
\providecommand{\bibinfo}[2]{#2}
\ifx\xfnm\relax \def\xfnm[#1]{\unskip,\space#1}\fi
\bibitem[{Bai(2005)}]{bai2005identification}
\bibinfo{author}{Bai, E.W.}, \bibinfo{year}{2005}.
\newblock \bibinfo{title}{Identification of nonlinear additive {FIR} systems}.
\newblock \bibinfo{journal}{Automatica} \bibinfo{volume}{41}, \bibinfo{pages}{1247--1253}.
\bibitem[{Bai and Chan(2008)}]{bai2008identification}
\bibinfo{author}{Bai, E.W.}, \bibinfo{author}{Chan, K.S.}, \bibinfo{year}{2008}.
\newblock \bibinfo{title}{Identification of an additive nonlinear system and its applications in generalized {H}ammerstein models}.
\newblock \bibinfo{journal}{Automatica} \bibinfo{volume}{44}, \bibinfo{pages}{430--436}.
\bibitem[{Billingsley(1995)}]{billingsley1995probability}
\bibinfo{author}{Billingsley, P.}, \bibinfo{year}{1995}.
\newblock \bibinfo{title}{Probability and Measure, \textnormal{3rd Edition}}.
\newblock \bibinfo{publisher}{John Wiley \& Sons}.
\bibitem[{Boeren et~al.(2018)Boeren, Blanken, Bruijnen and Oomen}]{boeren2018optimal}
\bibinfo{author}{Boeren, F.}, \bibinfo{author}{Blanken, L.}, \bibinfo{author}{Bruijnen, D.}, \bibinfo{author}{Oomen, T.}, \bibinfo{year}{2018}.
\newblock \bibinfo{title}{Optimal estimation of rational feedforward control via instrumental variables: {W}ith application to a wafer stage}.
\newblock \bibinfo{journal}{Asian Journal of Control} \bibinfo{volume}{20}, \bibinfo{pages}{975--992}.
\bibitem[{Bruce et~al.(2020)Bruce, Goel and Bernstein}]{bruceConvergenceConsistencyRecursive2020}
\bibinfo{author}{Bruce, A.L.}, \bibinfo{author}{Goel, A.}, \bibinfo{author}{Bernstein, D.S.}, \bibinfo{year}{2020}.
\newblock \bibinfo{title}{Convergence and consistency of recursive least squares with variable-rate forgetting}.
\newblock \bibinfo{journal}{Automatica} \bibinfo{volume}{119}, \bibinfo{pages}{109052}.
\bibitem[{Butcher et~al.(2008)Butcher, Karimi and Longchamp}]{butcher2008consistency}
\bibinfo{author}{Butcher, M.}, \bibinfo{author}{Karimi, A.}, \bibinfo{author}{Longchamp, R.}, \bibinfo{year}{2008}.
\newblock \bibinfo{title}{On the consistency of certain identification methods for linear parameter varying systems}.
\newblock \bibinfo{journal}{IFAC Proceedings Volumes} \bibinfo{volume}{41}, \bibinfo{pages}{4018--4023}.
\bibitem[{Classens et~al.(2021)Classens, Heemels and Oomen}]{classens2021fault}
\bibinfo{author}{Classens, K.}, \bibinfo{author}{Heemels, W.P.M.H.}, \bibinfo{author}{Oomen, T.}, \bibinfo{year}{2021}.
\newblock \bibinfo{title}{A closed-loop perspective on fault detection for precision motion control: With application to an overactuated system}, in: \bibinfo{booktitle}{2021 {IEEE} {I}nternational {C}onference on {M}echatronics ({ICM})}, \bibinfo{organization}{IEEE}. pp. \bibinfo{pages}{1--6}.
\bibitem[{Classens et~al.(2022)Classens, Mostard, van~de Wijdeven, Heemels and Oomen}]{classens2022fault}
\bibinfo{author}{Classens, K.}, \bibinfo{author}{Mostard, M.}, \bibinfo{author}{van~de Wijdeven, J.}, \bibinfo{author}{Heemels, W.P.M.H.}, \bibinfo{author}{Oomen, T.}, \bibinfo{year}{2022}.
\newblock \bibinfo{title}{Fault detection for precision mechatronics: Online estimation of mechanical resonances}.
\newblock \bibinfo{journal}{IFAC-PapersOnLine} \bibinfo{volume}{55}, \bibinfo{pages}{746--751}.
\bibitem[{Classens et~al.(2023)Classens, van~de Wijdeven, Heemels and Oomen}]{classens2023Opportunities}
\bibinfo{author}{Classens, K.}, \bibinfo{author}{van~de Wijdeven, J.}, \bibinfo{author}{Heemels, W.P.M.H.}, \bibinfo{author}{Oomen, T.}, \bibinfo{year}{2023}.
\newblock \bibinfo{title}{Opportunities of digital twins for high-tech systems: From fault diagnosis and predictive maintenance to control reconfiguration}.
\newblock \bibinfo{journal}{Mikroniek} \bibinfo{volume}{63}, \bibinfo{pages}{5--12}.
\bibitem[{Cover and Thomas(2006)}]{cover2006elements}
\bibinfo{author}{Cover, T.M.}, \bibinfo{author}{Thomas, J.A.}, \bibinfo{year}{2006}.
\newblock \bibinfo{title}{{E}lements of {I}nformation {T}heory, \textnormal{2nd Edition}}.
\newblock \bibinfo{publisher}{John Wiley \& Sons}.
\bibitem[{Garnier(2015)}]{garnier2015direct}
\bibinfo{author}{Garnier, H.}, \bibinfo{year}{2015}.
\newblock \bibinfo{title}{Direct continuous-time approaches to system identification. {O}verview and benefits for practical applications}.
\newblock \bibinfo{journal}{{E}uropean {J}ournal of {C}ontrol} \bibinfo{volume}{24}, \bibinfo{pages}{50--62}.
\bibitem[{Garnier and Wang(2008)}]{garnier2008book}
\bibinfo{author}{Garnier, H.}, \bibinfo{author}{Wang, L.}, \bibinfo{year}{2008}.
\newblock \bibinfo{title}{Identification of Continuous-time Models from Sampled Data}.
\newblock \bibinfo{publisher}{Springer}.
\bibitem[{Garnier and Young(2014)}]{garnier2014advantages}
\bibinfo{author}{Garnier, H.}, \bibinfo{author}{Young, P.C.}, \bibinfo{year}{2014}.
\newblock \bibinfo{title}{The advantages of directly identifying continuous-time transfer function models in practical applications}.
\newblock \bibinfo{journal}{International Journal of Control} \bibinfo{volume}{87}, \bibinfo{pages}{1319--1338}.
\bibitem[{Gawronski(2004)}]{gawronski2004advanced}
\bibinfo{author}{Gawronski, W.K.}, \bibinfo{year}{2004}.
\newblock \bibinfo{title}{Advanced Structural Dynamics and Active Control of Structures}.
\newblock \bibinfo{publisher}{Springer}.
\bibitem[{Gilson et~al.(2008)Gilson, Garnier, Young and Van~den Hof}]{gilson2008instrumental}
\bibinfo{author}{Gilson, M.}, \bibinfo{author}{Garnier, H.}, \bibinfo{author}{Young, P.C.}, \bibinfo{author}{Van~den Hof, P.M.J.}, \bibinfo{year}{2008}.
\newblock \bibinfo{title}{Instrumental variable methods for closed-loop continuous-time model identification}, in: \bibinfo{booktitle}{\textnormal{H. Garnier and L. Wang (Eds.). }{I}dentification of {C}ontinuous-time {M}odels from {S}ampled {D}ata}. \bibinfo{publisher}{Springer}, pp. \bibinfo{pages}{133--160}.
\bibitem[{Gilson et~al.(2011)Gilson, Garnier, Young and Van~den Hof}]{gilson2011optimal}
\bibinfo{author}{Gilson, M.}, \bibinfo{author}{Garnier, H.}, \bibinfo{author}{Young, P.C.}, \bibinfo{author}{Van~den Hof, P.M.J.}, \bibinfo{year}{2011}.
\newblock \bibinfo{title}{Optimal instrumental variable method for closed-loop identification}.
\newblock \bibinfo{journal}{IET Control Theory \& Applications} \bibinfo{volume}{5}, \bibinfo{pages}{1147--1154}.
\bibitem[{Gonz{\'a}lez et~al.(2024)Gonz{\'a}lez, Classens, Rojas, Welsh and Oomen}]{gonzalez2024statistical}
\bibinfo{author}{Gonz{\'a}lez, R.A.}, \bibinfo{author}{Classens, K.}, \bibinfo{author}{Rojas, C.R.}, \bibinfo{author}{Welsh, J.S.}, \bibinfo{author}{Oomen, T.}, \bibinfo{year}{2024}.
\newblock \bibinfo{title}{Statistical analysis of block coordinate descent algorithms for linear continuous-time system identification}.
\newblock \bibinfo{journal}{IEEE Control Systems Letters} .
\bibitem[{Gonz{\'a}lez et~al.(2025)Gonz{\'a}lez, Classens, Rojas, Welsh and Oomen}]{gonzalez2023identification}
\bibinfo{author}{Gonz{\'a}lez, R.A.}, \bibinfo{author}{Classens, K.}, \bibinfo{author}{Rojas, C.R.}, \bibinfo{author}{Welsh, J.S.}, \bibinfo{author}{Oomen, T.}, \bibinfo{year}{2025}.
\newblock \bibinfo{title}{Identification of additive continuous-time systems in open and closed loop}.
\newblock \bibinfo{journal}{Automatica} \bibinfo{volume}{173, \text{Article} 112013}.
\bibitem[{Gonz\'alez et~al.(2025)Gonz\'alez, Pan, Rojas and Welsh}]{gonzalez2023closedloop}
\bibinfo{author}{Gonz\'alez, R.A.}, \bibinfo{author}{Pan, S.}, \bibinfo{author}{Rojas, C.R.}, \bibinfo{author}{Welsh, J.S.}, \bibinfo{year}{2025}.
\newblock \bibinfo{title}{Consistency analysis of refined instrumental variable methods for continuous-time system identification in closed-loop}.
\newblock \bibinfo{journal}{Automatica} \bibinfo{volume}{166, \text{Article} 111697}.
\bibitem[{Gonz{\'a}lez et~al.(2023)Gonz{\'a}lez, Rojas, Pan and Welsh}]{gonzalez2023parsimonious}
\bibinfo{author}{Gonz{\'a}lez, R.A.}, \bibinfo{author}{Rojas, C.R.}, \bibinfo{author}{Pan, S.}, \bibinfo{author}{Welsh, J.S.}, \bibinfo{year}{2023}.
\newblock \bibinfo{title}{Parsimonious identification of continuous-time systems: A block-coordinate descent approach}, in: \bibinfo{booktitle}{15th {IFAC} {W}orld {C}ongress, {Y}okohama, {J}apan}.
\bibitem[{Ha and Welsh(2014)}]{ha2014ensuring}
\bibinfo{author}{Ha, H.}, \bibinfo{author}{Welsh, J.S.}, \bibinfo{year}{2014}.
\newblock \bibinfo{title}{Ensuring stability in continuous time system identification instrumental variable method for over-parameterized models}, in: \bibinfo{booktitle}{53rd {IEEE} {C}onference on {D}ecision and {C}ontrol ({CDC})}, pp. \bibinfo{pages}{2597--2602}.
\bibitem[{H{\"a}rdle et~al.(2004)H{\"a}rdle, Huet, Mammen and Sperlich}]{hardle2004bootstrap}
\bibinfo{author}{H{\"a}rdle, W.}, \bibinfo{author}{Huet, S.}, \bibinfo{author}{Mammen, E.}, \bibinfo{author}{Sperlich, S.}, \bibinfo{year}{2004}.
\newblock \bibinfo{title}{Bootstrap inference in semiparametric generalized additive models}.
\newblock \bibinfo{journal}{Econometric Theory} \bibinfo{volume}{20}, \bibinfo{pages}{265--300}.
\bibitem[{Hastie and Tibshirani(1986)}]{hastie1986generalized}
\bibinfo{author}{Hastie, T.}, \bibinfo{author}{Tibshirani, R.}, \bibinfo{year}{1986}.
\newblock \bibinfo{title}{{Generalized additive models}}.
\newblock \bibinfo{journal}{Statistical Science} \bibinfo{volume}{1}, \bibinfo{pages}{297 -- 310}.
\bibitem[{Van~den Hof(1998)}]{van1998closed}
\bibinfo{author}{Van~den Hof, P.M.J.}, \bibinfo{year}{1998}.
\newblock \bibinfo{title}{Closed-loop issues in system identification}.
\newblock \bibinfo{journal}{Annual Reviews in Control} \bibinfo{volume}{22}, \bibinfo{pages}{173--186}.
\bibitem[{Horn and Johnson(2012)}]{Horn2012}
\bibinfo{author}{Horn, R.A.}, \bibinfo{author}{Johnson, C.R.}, \bibinfo{year}{2012}.
\newblock \bibinfo{title}{Matrix Analysis, \textnormal{2nd Edition}}.
\newblock \bibinfo{publisher}{Cambridge University Press}.
\bibitem[{Ljung(1999)}]{ljung1998system}
\bibinfo{author}{Ljung, L.}, \bibinfo{year}{1999}.
\newblock \bibinfo{title}{System {I}dentification: {T}heory for the {U}ser, \textnormal{2nd ed}}.
\newblock \bibinfo{publisher}{Prentice-Hall}.
\bibitem[{Ljung and S{\"o}derstr{\"o}m(1983)}]{ljung1983theory}
\bibinfo{author}{Ljung, L.}, \bibinfo{author}{S{\"o}derstr{\"o}m, T.}, \bibinfo{year}{1983}.
\newblock \bibinfo{title}{Theory and {P}ractice of {R}ecursive {I}dentification}.
\newblock \bibinfo{publisher}{MIT Press}.
\bibitem[{Luenberger and Ye(2008)}]{luenberger2008linear}
\bibinfo{author}{Luenberger, D.G.}, \bibinfo{author}{Ye, Y.}, \bibinfo{year}{2008}.
\newblock \bibinfo{title}{Linear and {N}onlinear {P}rogramming, \textnormal{3rd edition}}.
\newblock \bibinfo{publisher}{Springer}.
\bibitem[{Mooren et~al.(2023)Mooren, Witvoet and Oomen}]{mooren2023online}
\bibinfo{author}{Mooren, N.}, \bibinfo{author}{Witvoet, G.}, \bibinfo{author}{Oomen, T.}, \bibinfo{year}{2023}.
\newblock \bibinfo{title}{On-line instrumental variable-based feedforward tuning for non-resetting motion tasks}.
\newblock \bibinfo{journal}{International Journal of Robust and Nonlinear Control} , \bibinfo{pages}{1--19}.
\bibitem[{Oomen(2018)}]{oomen2018advanced}
\bibinfo{author}{Oomen, T.}, \bibinfo{year}{2018}.
\newblock \bibinfo{title}{Advanced motion control for precision mechatronics: Control, identification, and learning of complex systems}.
\newblock \bibinfo{journal}{IEEJ Journal of Industry Applications} \bibinfo{volume}{7}, \bibinfo{pages}{127--140}.
\bibitem[{Opsomer and Ruppert(1999)}]{opsomer1999root}
\bibinfo{author}{Opsomer, J.D.}, \bibinfo{author}{Ruppert, D.}, \bibinfo{year}{1999}.
\newblock \bibinfo{title}{A root-n consistent backfitting estimator for semiparametric additive modeling}.
\newblock \bibinfo{journal}{Journal of Computational and Graphical Statistics} \bibinfo{volume}{8}, \bibinfo{pages}{715--732}.
\bibitem[{Padilla et~al.(2019)Padilla, Garnier, Young, Chen and Yuz}]{padilla2019identification}
\bibinfo{author}{Padilla, A.}, \bibinfo{author}{Garnier, H.}, \bibinfo{author}{Young, P.C.}, \bibinfo{author}{Chen, F.}, \bibinfo{author}{Yuz, J.}, \bibinfo{year}{2019}.
\newblock \bibinfo{title}{Identification of continuous-time models with slowly time-varying parameters}.
\newblock \bibinfo{journal}{Control Engineering Practice} \bibinfo{volume}{93}, \bibinfo{pages}{104165}.
\bibitem[{Padilla et~al.(2017)Padilla, Garnier, Young and Yuz}]{padilla2017recursive}
\bibinfo{author}{Padilla, A.}, \bibinfo{author}{Garnier, H.}, \bibinfo{author}{Young, P.C.}, \bibinfo{author}{Yuz, J.}, \bibinfo{year}{2017}.
\newblock \bibinfo{title}{Recursive online {IV} method for identification of continuous-time slowly time-varying models in closed loop}.
\newblock \bibinfo{journal}{IFAC-PapersOnLine} \bibinfo{volume}{50}, \bibinfo{pages}{4008--4013}.
\bibitem[{Pan et~al.(2020a)Pan, Gonz{\'a}lez, Welsh and Rojas}]{pan2020consistency}
\bibinfo{author}{Pan, S.}, \bibinfo{author}{Gonz{\'a}lez, R.A.}, \bibinfo{author}{Welsh, J.S.}, \bibinfo{author}{Rojas, C.R.}, \bibinfo{year}{2020}a.
\newblock \bibinfo{title}{Consistency analysis of the {S}implified {R}efined {I}nstrumental {V}ariable method for {C}ontinuous-time systems}.
\newblock \bibinfo{journal}{Automatica} \bibinfo{volume}{113}.
\bibitem[{Pan et~al.(2021)Pan, Welsh and Fu}]{pan2021identification}
\bibinfo{author}{Pan, S.}, \bibinfo{author}{Welsh, J.S.}, \bibinfo{author}{Fu, M.}, \bibinfo{year}{2021}.
\newblock \bibinfo{title}{Identification of continuous-time linear time-varying systems with abrupt changes in parameters}.
\newblock \bibinfo{journal}{IFAC-PapersOnLine} \bibinfo{volume}{54}, \bibinfo{pages}{339--344}.
\bibitem[{Pan et~al.(2020b)Pan, Welsh, Gonz{\'a}lez and Rojas}]{pan2020efficiency}
\bibinfo{author}{Pan, S.}, \bibinfo{author}{Welsh, J.S.}, \bibinfo{author}{Gonz{\'a}lez, R.A.}, \bibinfo{author}{Rojas, C.R.}, \bibinfo{year}{2020}b.
\newblock \bibinfo{title}{Efficiency analysis of the {S}implified {R}efined {I}nstrumental {V}ariable method for {C}ontinuous-time systems}.
\newblock \bibinfo{journal}{Automatica} \bibinfo{volume}{121, \textnormal{Article 109196}}.
\bibitem[{Rugh(1996)}]{rugh1996linear}
\bibinfo{author}{Rugh, W.J.}, \bibinfo{year}{1996}.
\newblock \bibinfo{title}{Linear {S}ystem {T}heory}.
\newblock \bibinfo{publisher}{Prentice-Hall}.
\bibitem[{S{\"o}derstr{\"o}m and Stoica(1983)}]{soderstrom1983instrumental}
\bibinfo{author}{S{\"o}derstr{\"o}m, T.}, \bibinfo{author}{Stoica, P.}, \bibinfo{year}{1983}.
\newblock \bibinfo{title}{Instrumental {V}ariable {M}ethods for {S}ystem {I}dentification}.
\newblock \bibinfo{publisher}{Springer-Verlag, Berlin}.
\bibitem[{S{\"o}derstr{\"o}m and Stoica(1989)}]{soderstrom1989system}
\bibinfo{author}{S{\"o}derstr{\"o}m, T.}, \bibinfo{author}{Stoica, P.}, \bibinfo{year}{1989}.
\newblock \bibinfo{title}{System {I}dentification}.
\newblock \bibinfo{publisher}{Prentice-Hall}.
\bibitem[{T{\'o}th(2010)}]{toth2010modeling}
\bibinfo{author}{T{\'o}th, R.}, \bibinfo{year}{2010}.
\newblock \bibinfo{title}{Modeling and {I}dentification of {L}inear {P}arameter-{V}arying {S}ystems}.
\newblock \bibinfo{publisher}{Springer}.
\bibitem[{Voorhoeve et~al.(2020)Voorhoeve, Aangenent and Oomen}]{voorhoeve2020identifying}
\bibinfo{author}{Voorhoeve, R. de~Rozario, R.}, \bibinfo{author}{Aangenent, W.}, \bibinfo{author}{Oomen, T.}, \bibinfo{year}{2020}.
\newblock \bibinfo{title}{Identifying position-dependent mechanical systems: A modal approach applied to a flexible wafer stage}.
\newblock \bibinfo{journal}{IEEE Transactions on Control Systems Technology} \bibinfo{volume}{22}, \bibinfo{pages}{194--206}.
\bibitem[{Young(2011)}]{young2011recursive}
\bibinfo{author}{Young, P.C.}, \bibinfo{year}{2011}.
\newblock \bibinfo{title}{Recursive {E}stimation and {T}ime-series {A}nalysis: {A}n {I}ntroduction for the {S}tudent and {P}ractitioner}.
\newblock \bibinfo{publisher}{Springer}.
\bibitem[{Young(2015)}]{young2015refined}
\bibinfo{author}{Young, P.C.}, \bibinfo{year}{2015}.
\newblock \bibinfo{title}{Refined instrumental variable estimation: {M}aximum {L}ikelihood optimization of a unified {B}ox--{J}enkins model}.
\newblock \bibinfo{journal}{Automatica} \bibinfo{volume}{52}, \bibinfo{pages}{35--46}.
\bibitem[{Young and Jakeman(1980)}]{young1980refined}
\bibinfo{author}{Young, P.C.}, \bibinfo{author}{Jakeman, A.J.}, \bibinfo{year}{1980}.
\newblock \bibinfo{title}{Refined instrumental variable methods of recursive time-series analysis {P}art {III}. {E}xtensions}.
\newblock \bibinfo{journal}{International Journal of Control} \bibinfo{volume}{31}, \bibinfo{pages}{741--764}.

\end{thebibliography}

\section{Appendices}
\appendix

\section{Derivation of the refined instrumental variable iterations}
\label{app:Deriv}
In this appendix, it is shown that a stationary point of the following optimization problem is obtained from the refined instrumental variable method. Recall the optimization problem
\begin{align}
	\theta_i^{l+1} (t_N) &= \underset{\theta_i \in \Omega_i}{\arg \min} \sum_{k=1}^N \alpha_{N-k} \bigg[ 	\tilde{y}_i(t_k) - G_i(p,t,\theta_i) u(t_k) \bigg]^2,
	\label{appeq:opt2}
\end{align}
with 
\begin{equation}
	\tilde{y}_i(t_k) := y(t_k)- \sum_{j=1}^{i-1} G_{j}(p,t,\theta_j^{l+1}) u(t_k)-\sum_{j=i+1}^{K} G_{j}(p,t,\theta_j^l) u(t_k). \notag
\end{equation}
and define
	\begin{equation}
	\varepsilon_{i}(t_k,\theta_{i}) = \tilde{y}_{i}(t_k) - G_i(p,t,\theta_i) u(t_k).  \notag
	\end{equation}
By taking \Cref{rem:slowly_varying_params} into consideration, this residual is well approximated as
\begin{equation}
    \varepsilon_{i}(t_k,\theta_{i}) = \frac{1}{A_i(p,\theta_i)} \left( A_i(p,\theta_i) \tilde{y}_{i}(t_k) - B_i(p,\theta_i) u(t_{k}) \right), \notag
\end{equation}
and subsequently $A_i(p,\theta_i)$ and $B_i(p,\theta_i)$ gives
\begin{align}
    \varepsilon_{i}(t_k,\theta_{i}) &= \frac{1}{A_i(p,\theta_i)} \Big( a_{i,n_{i}}(t_{k})p^{n_{i}} \tilde{y}_{i}(t_k) + \ldots \notag \\ &+a_{i,1}(t_{k})p \tilde{y}_{i}(t_k) + \tilde{y}_{i}(t_k) - b_{i,m_{i}}(t_{k})p^{m_{i}} u(t_{k}) - \ldots \notag \\ 
    & - b_{i,1}(t_{k})p u(t_{k}) - b_{i,0}(t_{k}) u(t_{k}) \Big). \notag 
\end{align}
The filtered residual output, filtered regressor vector, and parameter vector are defined as
\begin{align}
\tilde{y}_{i,\textnormal{f}}(t_k,\theta_i) &=  \frac{1}{A_i (p,\theta_i)} \tilde{y}_{i}(t_k), \notag \\
    \varphi_{i,\textnormal{f}} (t_{k}) &= \frac{1}{A_i (p,\theta_i)} \Big[
       \begin{matrix} - p \tilde{y}_{i}(t_k) \;\; \dots \;\; -p^{n_{i}} \tilde{y}_{i}(t_k) \end{matrix} \notag \\ 
       & \quad \begin{matrix} u(t_{k}) \;\; p u(t_{k}) \;\; \dots \;\; p^{m_{i}}u(t_{k}) \end{matrix}
    \Big]^\top, \notag \\
	\theta_i (t_{k}) &= \Big[ \begin{matrix}
		a_{i,1}(t_{k}) \;\; \dots \;\; a_{i,n_i}(t_{k}) \end{matrix} \notag \\ 
		& \quad \begin{matrix} b_{i,0}(t_{k}) \;\; b_{i,1}(t_{k}) \;\; \dots \;\; b_{i,m_i}(t_{k})
	\end{matrix} \Big]^\top, \notag 
\end{align}
which allows to write $\varepsilon_{i}$ as
	\begin{equation}
	\label{appeq:linear}
	\varepsilon_{i}(t_k,\theta_{i}) = \tilde{y}_{i,\textnormal{f}}(t_k,\theta_i) - \varphi_{i,\textnormal{f}}^\top(t_k,\theta_i)\theta_{i}(t_{k}),
	\end{equation}
which is a pseudolinear regression problem in $\theta_i$ that admits the refined instrumental variable \citep{garnier2008book,young2011recursive} iterations. To this end, the parameter estimate of the previous SRIVC iterate $s$ is taken in the denominator polynomial $A_{i}$ which is used as prefilter for $\tilde{y}_{i,\textnormal{f}}$ and $\varphi_{i,\textnormal{f}}$. Then, substitution in the first order optimality condition of \eqref{appeq:opt2}, i.e.,
\begin{equation}
\sum_{k=1}^N \alpha_{N-k} \frac{\partial \varepsilon_{i}(t_k,\theta_{i})}{\partial \theta_{i} } \varepsilon_{i}(t_k,\theta_{i}) = 0, \notag
\end{equation} 
with the instrument vector $\hat{\varphi}_{i,\textnormal{f}} (t_k, \theta_{i})$ gives
\begin{equation}
\sum_{k=1}^N \alpha_{N-k} \hat{\varphi}_{i,\textnormal{f}} (t_k, \theta_{i}) \left( \tilde{y}_{i,\textnormal{f}}(t_k, \theta_{i}) - \varphi_{i,\textnormal{f}}^{\top} (t_k, \theta_{i}) \theta_{i} (t_k) \right) = 0. \notag
\end{equation} 
Hence, the iterations for a fixed block descent iteration $l+1$ are of the form
	\begin{align}
		\theta_{i,s+1}^{l+1} (t_N) &= \left[\sum_{k=1}^N \alpha_{N-k} \hat{\varphi}_{i,\textnormal{f}} (t_k, \theta_{i,s}^{l+1}) \varphi_{i,\textnormal{f}}^\top(t_k, \theta_{i,s}^{l+1})\right]^{-1} \notag \\
        &\times\left[\sum_{k=1}^N \alpha_{N-k} \hat{\varphi}_{i,\textnormal{f}}(t_k, \theta_{i,s}^{l+1}) \tilde{y}_{i,\textnormal{f}}(t_k, \theta_{i,s}^{l+1})\right],
        \label{appeq:srivc}
	\end{align}
with $s\geq  1$.

\section{Recursive computation of the SRIVC iterations}
\label{app:Deriv2}
In order to reformulate \eqref{eq:srivc} into its recursive form, define	
		\begin{equation}
		R_{i,s}^{l+1}(t_N) =  \sum_{k=1}^N \alpha_{N-k} \hat{\varphi}_{i,\textnormal{f}} (t_k, \theta_{i,s}^{l+1}) \varphi_{i,\textnormal{f}}^\top(t_k, \theta_{i,s}^{l+1}),
	\end{equation}
and
\begin{equation}
f_{i,s}^{l+1} (t_N) = \sum_{k=1}^N \alpha_{N-k} \hat{\varphi}_{i,\textnormal{f}}(t_k, \theta_{i,s}^{l+1}) \tilde{y}_{i,\textnormal{f}}(t_k, \theta_{i,s}^{l+1}).
\end{equation}
Using that $\alpha_k = \lambda^{k}$ and that $\alpha$ satisfies the multiplicative property $\alpha_{k+1} = \lambda \alpha_k$, with $\alpha_0=1$,
\begin{equation}
		\label{appeq:pn}
		R_{i,s}^{l+1}(t_k) = \lambda \bar{R}_{i}(t_{k-1}) + \hat{\varphi}_{i,\textnormal{f}} (t_k, \theta_{i,s}^{l+1}) \varphi_{i,\textnormal{f}}^\top(t_k, \theta_{i,s}^{l+1}),
	\end{equation}
	and
\begin{equation}
			\label{appeq:fn}
		f_{i,s}^{l+1} (t_k)  = \lambda \bar{f}_{i} (t_{k-1}) + \hat{\varphi}_{i,\textnormal{f}} (t_k, \theta_{i,s}^{l+1}) \tilde{y}_{i,\textnormal{f}}(t_k, \theta_{i,s}^{l+1}),
\end{equation}
where $\bar{R}_{i}(t_{k-1})$ and $\bar{f}_{i} (t_{k-1})$ denote the final value of the iterations at the previous time step $t_{k-1}$ (i.e., after iterating over the modes through $i$ and $l$, and after the iterations in $s$). Using that $\theta_{i,s+1}^{l+1} (t_N) = \left( R_{i,s}^{l+1} (t_N) \right)^{-1} f_{i,s}^{l+1} (t_N)$, and  $\bar{f}_{i} (t_{k-1}) = \bar{R}_{i}(t_{k-1}) \bar{\theta}_{i} (t_{k-1})$,
	\begin{align}
		 \theta_{i,s+1}^{l+1} (t_N) & = \left( R_{i,s}^{l+1} (t_N) \right)^{-1} \notag \\ &\times ( \lambda \bar{R}_{i} (t_{N-1}) \bar{\theta}_{i} (t_{N-1}) + \hat{\varphi}_{i,\textnormal{f}} (t_N, \theta_{i,s}^{l+1}) \tilde{y}_{i,\textnormal{f}}(t_N, \theta_{i,s}^{l+1}) ), \notag
	\end{align}
which after substitution of $\bar{R}_{i}(t_{k-1})$ from \eqref{appeq:pn} is written as
	\begin{align}
		\theta_{i,s+1}^{l+1} (t_N) & = \left( R_{i,s}^{l+1} (t_N) \right)^{-1} \big(\hat{\varphi}_{i,\textnormal{f}} (t_N, \theta_{i,s}^{l+1}) \tilde{y}_{i,\textnormal{f}}(t_N, \theta_{i,s}^{l+1})\notag \\ &+( R_{i,s}^{l+1}(t_N) - \hat{\varphi}_{i,\textnormal{f}} (t_N, \theta_{i,s}^{l+1}) \varphi_{i,\textnormal{f}}^\top(t_N, \theta_{i,s}^{l+1}) ) \bar{\theta}_{i} (t_{N-1})\big), \notag 
	\end{align}
and thus
	\begin{align}
 \begin{split}
		\theta_{i,s+1}^{l+1} (t_N) & = \bar{\theta}_{i} (t_{N-1}) + \left( R_{i,s}^{l+1} (t_N) \right)^{-1}\hat{\varphi}_{i,\textnormal{f}}(t_N, \theta_{i,s}^{l+1}) \\ & \times\left( \tilde{y}_{i,\textnormal{f}}(t_N, \theta_{i,s}^{l+1}) - \varphi_{i,\textnormal{f}}^\top(t_N, \theta_{i,s}^{l+1}) \bar{\theta}_{i} (t_{N-1}) \right). 
\label{appeq:dertheta}
 \end{split}
	\end{align}
In order to avoid inverting $R_{i,s}^{l+1} (t_N)$ at each time step, introduce $P_{i,s}^{l+1} (t_N) = ( R_{i,s}^{l+1} (t_N))^{-1}$. The Sherman–Morrison formula \citep[Chap. 0.7.4]{Horn2012}, a generalization of the matrix inversion lemma, states that $\left( A + u v^{\top} \right)^{-1} = A^{-1} - \frac{ A^{-1} u v^{\top} A^{-1} }{ 1 + v^{\top} A^{-1} u}$, provided the inverses exist, with $A$ invertible and square, and column vectors $u$ and $v$. Take $A = \lambda \bar{R}_{i}(t_{N-1})$, $u = \hat{\varphi}_{i,\textnormal{f}} (t_N, \theta_{i,s}^{l+1})$, and $v = \varphi_{i,\textnormal{f}}(t_N, \theta_{i,s}^{l+1})$. Then,
\begin{align}
&\left( R_{i,s}^{l+1} (t_N) \right)^{-1} = \frac{1}{\lambda} \left( \bar{R}_{i}(t_{N-1}) \right)^{-1} \notag \\ & -\frac{ \frac{1}{\lambda} \left( \bar{R}_{i}(t_{N-1}) \right)^{-1} \hat{\varphi}_{i,\textnormal{f}} (t_N, \theta_{i,s}^{l+1}) \varphi_{i,\textnormal{f}}^\top(t_N, \theta_{i,s}^{l+1}) \frac{1}{\lambda} \left( \bar{R}_{i}(t_{N-1}) \right)^{-1} }{1 + \varphi_{i,\textnormal{f}}^\top(t_N, \theta_{i,s}^{l+1}) \frac{1}{\lambda} \left( \bar{R}_{i}(t_{N-1}) \right)^{-1} \hat{\varphi}_{i,\textnormal{f}} (t_N, \theta_{i,s}^{l+1}) }, \notag
\end{align}
and thus with $\bar{P}_{i}(t_{N-1}) = \left( \bar{R}_{i}(t_{N-1}) \right)^{-1}$,
\begin{align}
\begin{split}
P_{i,s}^{l+1} (t_N) & = \frac{1}{\lambda} \bar{P}_{i}(t_{N-1}) \\ & - \frac{1}{\lambda} \frac{ \bar{P}_{i}(t_{N-1}) \hat{\varphi}_{i,\textnormal{f}} (t_N, \theta_{i,s}^{l+1}) \varphi_{i,\textnormal{f}}^\top(t_N, \theta_{i,s}^{l+1}) \bar{P}_{i}(t_{N-1}) }{\lambda + \varphi_{i,\textnormal{f}}^\top(t_N, \theta_{i,s}^{l+1}) \bar{P}_{i}(t_{N-1}) \hat{\varphi}_{i,\textnormal{f}} (t_N, \theta_{i,s}^{l+1}) }.
\label{appeq:derP}
\end{split}
\end{align}
Define $K_{i,s+1}^{l+1} (t_{N}) = P_{i,s}^{l+1} (t_N) \hat{\varphi}_{i,\textnormal{f}}(t_N, \theta_{i,s}^{l+1})$, or equivalently,
\begin{align}
\begin{split}
		K_{i,s+1}^{l+1} (t_{N}) &= \bar{P}_{i}(t_{N-1}) \hat{\varphi}_{i,\textnormal{f}}(t_N,\theta_{i,s}^{l+1}) \\ & \times \left( \lambda + {\varphi}^{\top}_{i,\textnormal{f}}(t_N,\theta_{i,s}^{l+1}) \bar{P}_{i}(t_{N-1})  \hat{\varphi}_{i,\textnormal{f}}(t_N,\theta_{i,s}^{l+1}) \right)^{-1}.
\label{appeq:derK_final}
\end{split}
	\end{align}
This expression allows us to write \eqref{appeq:derP} as
\begin{equation}
P_{i,s}^{l+1} (t_N) = \frac{1}{\lambda} \left( \bar{P}_{i}(t_{N-1}) -  K_{i,s+1}^{l+1} (t_{N}) \varphi_{i,\textnormal{f}}^\top(t_N, \theta_{i,s}^{l+1}) \bar{P}_{i}(t_{N-1}) \right), \notag 
\end{equation}
and \eqref{appeq:dertheta} as
\begin{align}
		\theta_{i,s+1}^{l+1} (t_N) & = \bar{\theta}_{i} (t_{N-1}) \notag \\ &+ K_{i,s+1}^{l+1} (t_{N}) \left( \tilde{y}_{i,\textnormal{f}}(t_N, \theta_{i,s}^{l+1}) - \varphi_{i,\textnormal{f}}^\top(t_N, \theta_{i,s}^{l+1}) \bar{\theta}_{i} (t_{N-1}) \right). \notag 
	\end{align}
Hence, \eqref{appeq:derK_final}, \eqref{appeq:derP}, and \eqref{appeq:dertheta}, form a recursive equivalent of \eqref{appeq:srivc}.

\section{Generic consistency proof with $\alpha_{k} = 1$ for all $k$}
\label{Appendix:Deriv3}
In this appendix, the proof of \Cref{thm:Consistency_theorem} is provided.

\begin{proof}
As the number of SRIVC iterations $s$ tends to infinity,
	\begin{align}
 \begin{split}
		\theta_{i,\infty}^{l+1} &= \frac{1}{N} \left[\sum_{k=1}^N \hat{\varphi}_{i,\textnormal{f}} (t_k, \theta_{i,\infty}^{l+1}) \varphi_{i,\textnormal{f}}^\top(t_k, \theta_{i,\infty}^{l+1})\right]^{-1} \\ &\times \frac{1}{N} \left[\sum_{k=1}^N \hat{\varphi}_{i,\textnormal{f}}(t_k, \theta_{i,\infty}^{l+1}) \tilde{y}_{i,\textnormal{f}}(t_k, \theta_{i,\infty}^{l+1})\right].
  \end{split}
	\end{align}
That is,
\begin{equation}
\label{appeq:thatis1}
 \frac{1}{N} \sum_{k=1}^N \hat{\varphi}_{i,\textnormal{f}} (t_k, \theta_{i,\infty}^{l+1})\left[ \tilde{y}_{i,\textnormal{f}}(t_k, \theta_{i,\infty}^{l+1}) - \varphi_{i,\textnormal{f}}^\top(t_k, \theta_{i,\infty}^{l+1}) \theta_{i,\infty}^{l+1} \right] = 0,
\end{equation}
where
	\begin{align}
		&\tilde{y}_{i,\textnormal{f}}(t_k, \theta_{i,\infty}^{l+1}) - \varphi_{i,\textnormal{f}}^\top(t_k, \theta_{i,\infty}^{l+1}) \theta_{i,\infty}^{l+1} \notag \\
  \begin{split}
  &= \frac{1}{A_i (t_k, \theta_{i,\infty}^{l+1})} \tilde{y}_{i}(t_k) + \frac{p a_{i,\infty,1}^{l+1} + \ldots + p^{n_i} a_{i,\infty,n_{i}}^{l+1}  }{A_i (t_k, \theta_{i,\infty}^{l+1})} \tilde{y}_{i}(t_k)  \\
  &- \frac{b_{i,\infty,0}^{l+1} + p b_{i,\infty,1}^{l+1} + \ldots + p^{m_i} b_{i,\infty,m_{i}}^{l+1}  }{A_i (t_k, \theta_{i,\infty}^{l+1})} u(t_k)
   \end{split} \\
 &= \tilde{y}_{i}(t_k) - \frac{ B_i (t_k, \theta_{i,\infty}^{l+1}) }{A_i (t_k, \theta_{i,\infty}^{l+1})} u(t_k) \\
 \begin{split}
  &= y(t_k)- \sum_{j=1}^{i-1} G_{j}(p,t_k,\theta_j^{l+1}) u(t_k) -\sum_{j=i+1}^{K} G_{j}(p,t_k,\theta_j^l) u(t_k) \\ &- \frac{ B_i (t_k, \theta_{i,\infty}^{l+1}) }{A_i (t_k, \theta_{i,\infty}^{l+1})} u(t_k).
  \end{split}
	\end{align}
To simplify the notation, in the subsequent analysis, the $t_k$ dependency on the transfer functions is omitted. Denoting the parameter vector $\theta_{i,\infty}^{\infty}$ as the limit of $\theta_{i,\infty}^{l}$ when the number of descent iterations tends to infinity, then, as $l\to\infty$, 
	\begin{align}
 \begin{split}
		\tilde{y}_{i,\textnormal{f}}(t_k,\theta_{i,\infty}^{\infty}) &- \varphi_{i,\textnormal{f}}^\top(t_k,\theta_{i,\infty}^{\infty}) \theta_{i,\infty}^{\infty} \\
\label{appeq:thatis2}
&= y(t_k)-\sum_{j=1}^{K} G_{j} (p,\theta_{j,\infty}^{\infty}) u(t_k).
\end{split}
	\end{align}
On the other hand, note that
\begin{equation}
\label{appeq:thatis3}
\lim_{N\rightarrow \infty} \frac{1}{N}  \sum_{k=1}^N \hat{\varphi}_{i,\textnormal{f}} (t_k, \theta_{i,\infty}^{\infty}) v(t_k) = \bar{\mathbb{E}}\left\{\hat{\varphi}_{i,\textnormal{f}} (t_k, \theta_{i,\infty}^{\infty}) v(t_k)\right\}=0,
\end{equation}
where the first equality is due to the ergodicity property of the input and noise signals (see Theorem 2B.1 of \citet{butcher2008consistency} for more details), and the second equality holds since $v$ and $\hat{\varphi}_{i,\textnormal{f}}$ are uncorrelated due to Assumption \ref{ass:assumption4}. So, for $i=1,\dots, K$, combining \eqref{appeq:thatis1}, \eqref{appeq:thatis2}, \eqref{eq:systemadditive_modelset} and \eqref{appeq:thatis3} leads to  
	\begin{equation}
 \begin{split}
 \label{appeq:equivalentto}
	\lim_{N\to\infty}\frac{1}{N} \sum_{j=1}^K & \sum_{k=1}^N  \hat{\varphi}_{i,\textnormal{f}} (t_k, \theta_{i,\infty}^{\infty}) \\ &\times \left[ G_{j}^* (p) - G_{j} (p,\theta_{j,\infty}^{\infty}) \right]  u(t_k) = 0.
 \end{split}
	\end{equation}
Now,
\begin{align}
G_{j}^* (p) - G_{j} (p,\theta_{j,\infty}^{\infty}) &= \frac{B_j^* (p)}{A_j^* (p)} - \frac{ B_j (p, \theta_{j,\infty}^{\infty}) }{A_j (p, \theta_{j,\infty}^{\infty})} \notag \\
&= \frac{B_j^* (p) A_j (p, \theta_{j,\infty}^{\infty}) - A_j^* (p) B_j (p, \theta_{j,\infty}^{\infty}) }{A_j^* (p) A_j (p, \theta_{j,\infty}^{\infty}) } \notag \\
&= \frac{ \begin{bmatrix} 1, & p, & \ldots, & p^{n_j+m_j} \end{bmatrix} }{A_j^* (p) A_j (p, \theta_{j,\infty}^{\infty}) } h_j,
\end{align}
where $h_j\in \mathbb{R}^{n_j+m_j+1}$ is the vector that contains the coefficients of $B_j^* (p) A_j (p, \theta_{j,\infty}^{\infty}) - A_j^* (p) B_j (p, \theta_{j,\infty}^{\infty})$ in increasing order of degree. With regards to the instrument vector, the following decomposition holds (see, e.g., Eq. (16) of \citet{pan2020consistency})
\begin{equation}
        \hat{\varphi}_{i,\textnormal{f}} (t_k, \theta_{i,\infty}^{\infty}) = S(-B_i, A_i) \frac{1}{A_i^2 (p, \theta_{i,\infty}^{\infty})} U_i(t_k),
\end{equation}
where $U_i(t_k) = \begin{bmatrix} 1 & p & \ldots & p^{n_i+m_i} \end{bmatrix}^{\top} u(t_k)$, and where $S(-B_i,A_i)$ is a Sylvester matrix formed by the coefficients of the polynomials $-B_i (p, \theta_{i,\infty}^{\infty})$ and $A_i (p, \theta_{i,\infty}^{\infty})$, similar to Eq. (13) of \citet{pan2020consistency}. Thus, \eqref{appeq:equivalentto} is equivalent to
	\begin{align}
 \begin{split}
		\sum_{j=1}^K \lim_{N \rightarrow \infty} \frac{1}{N} \sum_{k=1}^N & S(-B_i, A_i) \frac{1}{A_i^2 (t_k, \theta_{i,\infty}^{\infty})} U_i(t_k) \\ & \times \frac{1}{A_j^* (p) A_j (p, \theta_{j,\infty}^{\infty}) } U^\top_j(t_k) h_j = 0,
 \end{split}
	\end{align}
for $i = 1, \ldots, K$. After applying Lemma \ref{lem:technical} in \ref{app:technicallemmas} and the ergodicity result in Theorem 2B.1 of \citet{butcher2008consistency}, which hold since the system and model denominators describe uniformly stable transfer functions and the input is assumed quasi-stationary, the following is obtained.
\begin{align}
\begin{split}
S(-\bar{B}_i, \bar{A}_i)\sum_{j=1}^K \bar{\mathbb{E}} & \bigg\{\frac{1}{A_i^2 (t_k, \theta_{i,\infty}^{\infty})} U_i(t_k) \\ & \times
\label{kconditions}
 \frac{1}{A_j^* (p) A_j (p, \theta_{j,\infty}^{\infty}) } U^{\top}_j(t_k) \bigg\} \bar{h}_j = 0,
\end{split}
\end{align}
for $i = 1, \ldots, K$, where $(\bar{A}_i, \bar{B}_i)=\lim_{N\to\infty}(A_i(p,\theta_{i,\infty}^\infty), B_i(p,\theta_{i,\infty}^\infty))$, and $\bar{h}_j = \lim_{N\to\infty}h_j$. The $K$ conditions in \eqref{kconditions} are jointly expressed as $\mathcal{S} \Phi_u \eta=0$, where $\mathcal{S}$ is a block-diagonal matrix whose block-diagonal is comprised by $\{S(-\bar{B}_i, \bar{A}_i)\}_{k=1}^K$, while $\eta=[\bar{h}_1^\top,\dots,\bar{h}_K^\top]^\top$, and $\Phi_u$ has block entries given by
\begin{equation}
\Phi_u^{i,j} = \bar{\mathbb{E}} \left\{  \frac{1}{A_i^2 (t_k, \theta_{j,\infty}^{\infty})} U_i(t_k) \frac{1}{A_j^* (p) A_j (t_k, \theta_{j,\infty}^{\infty}) } U_j(t_k) \right\}. \notag
\end{equation}
Assumption \ref{ass:assumption3} regarding the coprimeness of $\bar{A}_i$ and $\bar{B}_i$ implies that $\mathcal{S}$ is nonsingular, as shown in Lemma A3.1 of \citet{soderstrom1983instrumental}. Moreover, Lemma \ref{lem:technical2} in \ref{app:technicallemmas} applied over each entry of $\Phi_u$ indicates that the non-singularity of $\Phi_u$ can be studied directly from the cross covariance of the inputs filtered by the LTI systems obtained at the limit as $N\to \infty$. Thus, following the same reasoning as in Lemma 3 of \citet{gonzalez2023identification}, Assumptions \ref{ass:assumption1} to \ref{ass:assumption4} imply the generic nonsingularity of $\Phi_u$ with respect to the converging system and model denominator parameters. Consequently, $\eta=0$, i.e., $\lim_{N\to\infty} h_j = 0$ for all $i=1,\dots,K$, as desired.
\end{proof}

\subsection{Technical Lemmas}
\label{app:technicallemmas}
\begin{lemma}
\label{lem:technical}
    If the matrix $\mathcal{O}_k$ has a bounded norm for all $k$ with probability 1, then, with probability 1,
    \begin{equation}
\label{appeq:technicallemma}
        \lim_{N\to\infty}\frac{1}{N} \sum_{k=1}^N S_k \mathcal{O}_k h_k = \bar{S} \lim_{N\to\infty}\frac{1}{N} \sum_{k=1}^N \mathcal{O}_k \bar{h},       
    \end{equation}
where $S_k$ and $h_k$ are matrices and vectors of appropriate dimensions, such that $\lim_{k\to\infty}S_k = \bar{S}$ and $\lim_{k\to\infty}h_k = \bar{h}$.
\end{lemma}
\begin{proof}
    Note that the following inequalities hold:
    \begin{align}
        \frac{1}{N} \left\|\sum_{k=1}^N (S_k-\bar{S}) \mathcal{O}_k h_k \right\| &\leq \frac{1}{N} \sum_{k=1}^N \|S_k-\bar{S}\| \|\mathcal{O}_k h_k\| \notag \\
        \label{appeq:cesaro}
        &\leq \frac{M_1}{N}\sum_{k=1}^N \|S_k-\bar{S}\|,
    \end{align}
    where $M_1 = \sup_{k}\|\mathcal{O}_k h_k\|<\infty$ with probability 1, since $\mathcal{O}_k$ and $h_k$ are bounded. Since $\|S_k-\bar{S}\|\to 0$ as $k\to \infty$, then the Ces\`aro mean in \eqref{appeq:cesaro} also goes to zero due to Theorem 4.2.3 of \citet{cover2006elements}. By the same reasoning, it follows that
    \begin{align}
        \frac{1}{N} \left\|\sum_{k=1}^N \bar{S} \mathcal{O}_k (h_k-\bar{h}) \right\| &\leq \frac{1}{N} \sum_{k=1}^N \|\bar{S}\mathcal{O}_k\|\|h_k-\bar{h}\| \notag \\ &\leq \frac{M_2}{N} \sum_{k=1}^N \|h_k-\bar{h}\| \xrightarrow[]{N\to \infty} 0. \notag
    \end{align}
    Thus, the result in \eqref{appeq:technicallemma} follows from decomposing $S_k \mathcal{O}_k h_k$ as 
\begin{equation}
    S_k \mathcal{O}_k h_k = (S_k-\bar{S}) \mathcal{O}_k h_k + \bar{S} \mathcal{O}_k (h_k-\bar{h}) + \bar{S} \mathcal{O}_k \bar{h}, \notag
\end{equation}
    summing over $k$, dividing over $N$, and computing the limit as $N\to\infty$.
\end{proof}

\begin{lemma}
\label{lem:technical2}
    Let
    \begin{equation}
    \label{appeq:wandv}
        w(t) = \sum_{k=0}^\infty \alpha_t(k)e(t-k), \quad v(t) = \sum_{k=0}^\infty \beta_t(k)e(t-k),
    \end{equation}
where the families of filters $\alpha_t(k),\beta_t(k)$, $t=1,2,\dots$ are uniformly stable, and $\{e(t)\}$ is a quasi-stationary signal. In addition, let $\bar{w}(t)$ and $\bar{v}(t)$ be defined as in $\eqref{appeq:wandv}$ but for the LTI filters $\bar{\alpha}(k)$ and $\bar{\beta}(k)$ respectively, where $\alpha_t(k)\to \bar{\alpha}(k)$ and $\beta_t(k)\to \bar{\beta}(k)$ for every $k$ as $t\to\infty$. Then,
\begin{equation}
\label{appeq:toprovelemma}
    \left|\bar{\mathbb{E}}\{w(t)v(t)\}-\bar{\mathbb{E}}\{\bar{w}(t)\bar{v}(t)\}\right| \to 0 \quad \textnormal{as } N\to\infty.
\end{equation}
\end{lemma}
\begin{proof}
    The following inequalities hold:
    \begin{align}
        &\left|\bar{\mathbb{E}}\{w(t)v(t)\}-\bar{\mathbb{E}}\{\bar{w}(t)\bar{v}(t)\}\right| \notag \\
        \label{appeq:ineq1}
        \begin{split}= \Bigg|\frac{1}{N}\sum_{t=1}^N  \sum_{k=0}^\infty \sum_{\ell=0}^\infty &\mathbb{E}\{e(t-k)e(t-\ell)\} \\ &\times\big( \alpha_t(k)\beta_t(\ell)-\bar{\alpha}(k)\bar{\beta}(\ell)\big) \Bigg| \\
                \end{split}\\
                \label{appeq:ineq2}
\begin{split} \leq \frac{C}{N} \sum_{t=1}^N  \sum_{k=0}^\infty  \sum_{\ell=0}^\infty  \big| & \big( \alpha_t(k) (\beta_t(\ell)-\bar{\beta}(\ell)) \\ &+\bar{\beta}(\ell)(\alpha_t(k)-\bar{\alpha}(k)\big) \big| \\
        \end{split} \\
        \begin{split}
        &\leq \frac{C}{N} \sup_{t\in\mathbb{N}} \sum_{k=0}^\infty |\alpha_t(k)| \sum_{t=1}^N   \sum_{\ell=0}^\infty \left|\beta_t(\ell)-\bar{\beta}(\ell)\right|  \\
        \label{appeq:ineq3}
        &\qquad \qquad \qquad \:\:\:\: + \frac{C}{N} \sum_{\ell=0}^\infty |\bar{\beta}(\ell)| \sum_{t=1}^N \sum_{k=0}^\infty \left|\alpha_t(k)-\bar{\alpha}(k)\right|,
        \end{split}
    \end{align}
where \eqref{appeq:ineq1} holds by definition, \eqref{appeq:ineq2} uses the quasi-stationary property $|\mathbb{E}\{e(t-k)e(t-\ell)\}|\leq C$ for any indices $t, k$ and $\ell$, and \eqref{appeq:ineq3} is due to the triangle inequality. Since $\alpha_t(k)$ and $\beta_t(\ell)$ describe uniformly stable families of filters, the two series $\sup_{t\in\mathbb{N}} \sum_{k=0}^\infty |\alpha_t(k)|$ and $\sum_{k=0}^\infty |\bar{\beta}(k)|$ converge. Thus, to prove \eqref{appeq:toprovelemma} we require to show that 
\begin{equation}
    \label{appeq:showthat}
    \frac{1}{N}\sum_{t=1}^N \sum_{k=0}^\infty \left|\alpha_t(k)-\bar{\alpha}(k)\right| \to 0 \quad \textnormal{as }N\to\infty,
\end{equation}
and similarly with $\beta_t(k)$. Due to the pointwise convergence, $\lim_{t\to\infty}|\alpha_t(k)-\bar{\alpha}(k)|=0$, and due to the uniform stability of $\alpha_t(k)$, 
\begin{equation}
    \sum_{k=0}^\infty |\alpha_t(k)-\bar{\alpha}(k)|\leq 2 \sum_{k=0}^\infty \sup_{t\in \mathbb{N}} |\alpha_t(k)|<\infty. \notag
\end{equation}
Thus, by the Weierstrass M-test \citep[Theorem A28]{billingsley1995probability},
\begin{equation}
    \lim_{t\to\infty} \sum_{k=0}^\infty |\alpha_t(k)-\bar{\alpha}(k)| = \sum_{k=0}^\infty \lim_{t\to\infty}|\alpha_t(k)-\bar{\alpha}(k)| =0. \notag
\end{equation}
Hence, \eqref{appeq:showthat} can be interpreted as a Ces\`aro mean with sequence in $t$ given by $\sum_{k=0}^\infty |\alpha_t(k)-\bar{\alpha}(k)|$, which goes to zero as $t$ tends to infinity. Thus, by Theorem 4.2.3 of \citet{cover2006elements}, \eqref{appeq:showthat} holds. The same derivation holds for $\beta_t(\ell)$, which implies that the right hand side of \eqref{appeq:ineq3} tends to zero as $N$ tends to infinity, thus proving \eqref{appeq:toprovelemma}.
\end{proof}

\end{document}